\documentclass[aps, prd,  superscriptaddress, showpacs, longbibliography, floatfix, nofootinbib]{revtex4-2}

\usepackage{bm, amsmath, amssymb, relsize, amsfonts, mathrsfs, multirow, braket, siunitx, color, booktabs, arydshln}
\usepackage{graphicx}
\graphicspath{{./figures/}}

\DeclareSIUnit \parsec {pc}

\usepackage[dvipsnames]{xcolor}
\usepackage[unicode]{hyperref}
\hypersetup{
  colorlinks=true,
  citecolor=RoyalBlue,
  linkcolor=blue,
  urlcolor=blue
}

\usepackage[caption=false]{subfig}
\usepackage{array}
\usepackage{booktabs}
\usepackage{multirow}
\usepackage{cancel}
\usepackage{orcidlink}
\usepackage{graphicx}
\usepackage{slashed}
\usepackage{dcolumn}
\usepackage{bm}
\usepackage{paralist}
\usepackage{epsfig}
\usepackage{placeins}
\usepackage{mathrsfs}
\usepackage{comment}
\usepackage{color, colortbl}
\usepackage[inline]{enumitem}
\usepackage{soul}
\usepackage{adjustbox}
\usepackage{amssymb}
\usepackage{caption}
\usepackage{subcaption}
\usepackage{float}
\usepackage{cancel}
\usepackage{verbatim}
\usepackage{amsmath}
\usepackage[toc,page]{appendix}
\usepackage{graphicx}

\DeclareMathAlphabet{\mathpzc}{OT1}{pzc}{m}{it}
	
\definecolor{LightCyan}{rgb}{0.88,1,1}
\definecolor{lightgray}{gray}{0.9}

\def \IITGn     {Department of Physics, Indian Institute of Technology Gandhinagar, Gujarat 382055, India.\vspace*{4pt}}

\begin{document}

\title{ILC Phenomenology of the $Z_3$ symmetric Type-Z Three Higgs Doublet Model}

\author{\textsc{Baradhwaj Coleppa}\orcidlink{0000-0002-8761-3138}}
\email{baradhwaj@iitgn.ac.in }
\affiliation{\IITGn}

\author{\textsc{Akshat Khanna}\orcidlink{0000-0002-2322-5929}\vspace*{7pt}} 
\email{khanna\_akshat@iitgn.ac.in}
\affiliation{\IITGn}

\author{\textsc{Gokul B. Krishna}\orcidlink{0000-0001-7083-4308}}
\email{gokulb@iitgn.ac.in}
\affiliation{\IITGn}

\begin{abstract}
The Three-Higgs-Doublet Model (3HDM) extends the Standard Model by introducing two additional scalar doublets, leading to a rich spectrum of new particles: three neutral CP-even Higgs bosons ($h_1$, $H_2$, $H_3$), two neutral CP-odd Higgs bosons ($A_2$, $A_3$), and two charged Higgs bosons ($H_2^+$, $H_3^+$). In this work, we present a phenomenological study of the 3HDM at the future International Linear Collider (ILC) with a center-of-mass energy of $\sqrt{s} = 1000\,\text{GeV}$. Applying a comprehensive set of theoretical and experimental constraints, we identify promising new physics signals with sufficiently large production cross-sections. Our analysis shows that $e^+e^- \to H_2 A_2$, $e^+e^- \to H_2 H_2 Z$, $A_2 A_2 Z$, $H_2 H_2^{\pm} W^{\mp}$, $A_2 H_2^{\pm} W^{\mp}$ and $H_1 H_2 A_2$ are among the most sensitive channels to probe this extended Higgs sector. We demonstrate that a future ILC would offer a powerful platform to test these interactions and discover these heavier Higgs bosons thus providing evidence of physics beyond the Standard Model.

\end{abstract}
\maketitle


\section{Introduction}
\label{sec:intro}
The Standard Model (SM) of particle physics has been the most successful theory so far, verified and established by past and current experimental setups. The final validation of the SM came with the discovery of the Higgs boson at the Large Hadron Collider (LHC) in 2012 \cite{ATLAS:2012yve,CMS:2012qbp}. Although the SM has been the cornerstone of particle physics, it has long been known that it needs to be supplanted with new dynamics at the TeV scale owing to many theoretical and experimental reasons - the evidence of dark matter and neutrino mass are two of the most pressing theoretical issues that do not find a resolution within the SM. One of the many ways the  theoretical framework of the SM can be extended is by adding additional scalar sectors, such as singlet \cite{Barger:2008jx,Tavartkiladze:2022pzf} and doublet extensions \cite{Branco:2011iw}, and their generalization to multi-Higgs doublet models \cite{Bento:2017eti,Bento:2018fmy}. Along these lines, the 3HDM~\cite{Batra:2025amk,Keus:2013hya, Das:2019yad, Boto:2023nyi, PhysRevLett.37.657, Ivanov:2017dad, Cordero:2017owj, Dey:2023exa, Ivanov:2012fp, Ivanov:2012ry} has also been a long-standing source of investigation. These models have been key in addressing unresolved questions, such as providing new sources of CP violation which are crucial for explaining baryogenesis \cite{Morrissey:2012db}, enabling strong first-order electroweak phase transitions, a necessary condition for successful electroweak baryogenesis \cite{Cline:2006ts}. In certain realizations, the extended scalar sector can also offer viable dark matter candidates through neutral scalars \cite{Barbieri:2006dq}. Beyond cosmological implications, such models predict a richer Higgs spectrum with distinct collider signatures, offering promising discovery opportunities at current and future colliders like the LHC and International Linear Collider (ILC)~\cite{LHeCStudyGroup:2012zhm}. While models with two Higgs doublets have been extensively studied \cite{Branco:2011iw}, the 3HDM offers an even richer framework for addressing fundamental questions beyond the Standard Model. Furthermore, the additional scalar degrees of freedom enable more flexible Yukawa structures, providing potential explanations for the observed patterns of fermion masses and mixing \cite{Ogreid:2017alh}. From a phenomenological perspective, the extended scalar spectrum enables a variety of novel collider signatures including multiple charged and neutral Higgs bosons that can be probed at future experiments such as the ILC or the Future Circular Collider (FCC-ee)~\cite{Adolphsen:2022ibf}. For a complete review of BSM scenarios with extended scalar sectors, the reader is invited to consult Ref.~\cite{Ivanov:2017dad}.  

In this paper, we explore the phenomenological aspects of the $Z_3$ symmetric Type Z 3HDM from the standpoint of the future ILC experiment. As the ILC global project proposal \cite{Bambade:2019fyw,Barklow:2015tja} continues to evolve and gain traction within the community, it is timely and important to investigate the avenues through which various Beyond Standard Model (BSM) scenarios could manifest in its setup. With this motivation, we investigate the phenomenology of the heavy Higgs bosons in the 3HDM at the ILC operating at a center-of-mass energy of $\sqrt{s} =$ 1000 GeV \cite {Adolphsen:2013kya}. We focus on channels that are expected to have reasonable cross sections while operating with benchmark points allowed by all existing theoretical and experimental constraints. The paper is organized as follows: in~\autoref{sec:Model}, we provide a brief overview of the 3HDM concentraing on the scalar structure,  and list the theoretical and experimental constraints on the model. The phenomenological analysis of the various relevant channels at the ILC is presented in~\autoref{sec:Pheno}, and we present our conclusions in~\autoref{sec:Conclusions}.

\section{Model and Constraints}
\label{sec:Model}

In this section, we present a concise overview of the model, focusing on the scalar and Yukawa sectors. We outline the structure of the scalar potential, detail the transformations between gauge and mass eigenstates, and comment on the Yukawa sector. We then discuss the alignment limit conditions, followed by a summary of the theoretical and experimental constraints imposed to identify viable benchmark points for the subsequent phenomenological analysis.

\subsection{Scalar Sector}
The Three Higgs Doublet model extends SM scalar sector by introducing two additional $SU(2)_L$ doublets represented as 
\begin{equation}
		\Phi_k = \begin{pmatrix}
			\phi_k^+ \\ \frac{v_k+p_k+in_k}{\sqrt{2}}
		\end{pmatrix},
  \label{eq:phi_form}
\end{equation}
where $k=1,2,3$. We consider a $Z_3$ symmetric potential \cite{PhysRevD.108.015020, Boto:2021qgu,Aranda:2019vda} under which the Higgs fields transform as 
\begin{equation}
     	\label{eq:phitrans}
     	\Phi_1 \rightarrow \omega \Phi_1, \; \; \Phi_2 \rightarrow \omega^2 \Phi_2, \;\textrm{and} \;\Phi_3 \rightarrow \Phi_3,
\end{equation}
where $\omega = e^{2\pi i/3}$ are the nontrivial cubic roots of unity. The most general $SU(2)_L \times U(1)_Y$ invariant potential preserving the $Z_3$ symmetry is given by
\begin{equation}
     	\label{eq:scalarpot}
     	\begin{split}
     		V & = m_{11}^2(\Phi_1^\dagger\Phi_1) + m_{22}^2(\Phi_2^\dagger\Phi_2) + m_{33}^2(\Phi_3^\dagger\Phi_3)
     		\\ & + \lambda_1(\Phi_1^\dagger\Phi_1)^2 + \lambda_2(\Phi_2^\dagger\Phi_2)^2 + \lambda_3(\Phi_3^\dagger\Phi_3)^2 \\ & + \lambda_{4}(\Phi_1^\dagger\Phi_1)(\Phi_2^\dagger\Phi_2) + \lambda_{5}(\Phi_1^\dagger\Phi_1)(\Phi_3^\dagger\Phi_3) + \lambda_{6}(\Phi_2^\dagger\Phi_2)(\Phi_3^\dagger\Phi_3) \\ & + \lambda_{7}(\Phi_1^\dagger\Phi_2)(\Phi_2^\dagger\Phi_1) + \lambda_{8}(\Phi_1^\dagger\Phi_3)(\Phi_3^\dagger\Phi_1) +  \lambda_{9}(\Phi_2^\dagger\Phi_3)(\Phi_3^\dagger\Phi_2) \\ & + [\lambda_{10}(\Phi_1^\dagger\Phi_2)(\Phi_1^\dagger\Phi_3) + \lambda_{11}(\Phi_1^\dagger\Phi_2)(\Phi_3^\dagger\Phi_2) + \lambda_{12}(\Phi_1^\dagger\Phi_3)(\Phi_2^\dagger\Phi_3) + h.c.]. \\ &
     	\end{split} 
\end{equation}
Here, in general $\lambda_{1,...,9}$ are real parameters, while $ \lambda_{10}, \lambda_{11}, \lambda_{12}$ can be complex. However, here we disregard the complex phases of the potential to prevent mixing between the CP-odd and CP-even states \cite{Batra:2025amk}. Consequently, the scalar potential is CP conserving, with no explicit CP violation. Further, in what follows, we also assume there is no spontaneous CP violation as well, i.e., we assume no relative phases in the vacuum expectation values are generated through minimization. The relevant mass terms for the CP-even Higgs are deduced from the scalar potential in~\autoref{eq:scalarpot}, and can be written as 
\begin{equation*}
		V_p^{mass} \supset \begin{pmatrix}
			p_1 & p_2 & p_3
		\end{pmatrix} \frac{\mathcal{M}^2_S}{2} \begin{pmatrix}
			p_1 \\ p_2 \\ p_3 
		\end{pmatrix},
\end{equation*}
As a real symmetric matrix, it can be diagonalized through an orthogonal transformation characterized by the matrix $O_\alpha$. The diagonalization condition is given by
\begin{equation}
    O_\alpha\mathcal{M}^2_SO_\alpha^T = \begin{pmatrix}
        m_{H 1}^2 & 0 & 0 \\ 0 & m_{H 2}^2 & 0 \\ 0 & 0 & m_{H 3}^2
    \end{pmatrix}
\end{equation}
The explicit relations of $\lambda_{1,2,..,6}$ can be expressed in terms of the masses of the three CP-even Higgs bosons and three mixing angles. The gauge eigenstates can be related to the mass eigenstates thus:
\begin{equation}
    \begin{split}
        H_1 & = c_{\alpha_2} c_{\alpha_1} p_1 + c_{\alpha_2} s_{\alpha_1} p_2 + s_{\alpha_2}p_3, \\
        H_2 & = -(c_{\alpha_3} s_{\alpha_1} + s_{\alpha_3} s_{\alpha_2} c_{\alpha_1})p_1 + (c_{\alpha_3} c_{\alpha_1} - s_{\alpha_3} s_{\alpha_2} s_{\alpha_1})p_2+(s_{\alpha_3} c_{\alpha_2})p_3,\,\textrm{and} \\
        H_3 & =  (s_{\alpha_3} s_{\alpha_1} - c_{\alpha_3} s_{\alpha_2} c_{\alpha_1})p_1 -  (s_{\alpha_3} c_{\alpha_1} + c_{\alpha_3} s_{\alpha_2} s_{\alpha_1})p_2 + (c_{\alpha_3} c_{\alpha_2}) p_3.
    \end{split}
\end{equation}
The mass terms for the charged Higgses can similarly be extracted from the scalar potential given in~\autoref{eq:scalarpot}:
\begin{equation*}
     	V_C^{mass} \supset \begin{pmatrix}
     		\phi_1^- & \phi_2^- & \phi_3^-
     	\end{pmatrix} \mathcal{M}^2_{\phi^{\pm}} \begin{pmatrix}
     		\phi_1^+ \\ \phi_2^+ \\ \phi_3^+ 
     	\end{pmatrix},
\end{equation*}
The diagonalization of this matrix is performed through two successive transformations, and the relation between the gauge and mass eigenstates is given by
\begin{equation*}
    \begin{pmatrix}
        \phi_1^\pm \\ \phi_2^\pm \\ \phi_3^\pm
    \end{pmatrix} = (O_{\gamma 2}.O_\beta)^\dagger \begin{pmatrix}
        G^\pm \\ H_2^\pm \\ H_3^\pm
    \end{pmatrix},
\end{equation*}
where
\begin{equation}
    \label{chargetransmat}
    (O_{\gamma 2}O_\beta)^\dagger = \begin{pmatrix}
        c_{\beta 1}c_{\beta 2} & -c_{\gamma 2 }s_{\beta 1} + c_{\beta 1}s_{\beta 2}s_{\gamma 2} & -c_{\beta 1}c_{\gamma 2}s_{\beta 2}-s_{\beta 1}s_{\gamma 2} \\
        c_{\beta 2}s_{\beta 1} & c_{\beta 1}c_{\gamma 2} + s_{\beta 1}s_{\beta 2}s_{\gamma 2} & -c_{\gamma 2}s_{\beta 1}s_{\beta 2}+c_{\beta 1}s_{\gamma 2} \\
        s_{\beta 2} & -c_{\beta 2}s_{\gamma 2} & c_{\beta 2}c_{\gamma 2} 
    \end{pmatrix}.
\end{equation}
The gauge eigenstates can be represented in terms of mass eigenstates as
\begin{equation}
    \label{chargetransrelat}
    \begin{split}
        G^\pm & = c_{\beta 1}c_{\beta 2} \phi_1^\pm + c_{\beta 2}s_{\beta 1} \phi_2^\pm + s_{\beta 2} \phi_3^\pm, \\
        H_2^\pm & = (-c_{\gamma 2 }s_{\beta 1} + c_{\beta 1}s_{\beta 2}s_{\gamma 2})\phi_1^\pm + (c_{\beta 1}c_{\gamma 2} + s_{\beta 1}s_{\beta 2}s_{\gamma 2})\phi_2^\pm + (-c_{\beta 2}s_{\gamma 2}) \phi_3^\pm,\,\textrm{and} \\
        H_3^\pm & = (-c_{\beta 1}c_{\gamma 2}s_{\beta 2}-s_{\beta 1}s_{\gamma 2})\phi_1^\pm + (-c_{\gamma 2}s_{\beta 1}s_{\beta 2}+c_{\beta 1}s_{\gamma 2})\phi_2^\pm + (c_{\beta 2}c_{\gamma 2}) \phi_3^\pm.
    \end{split}
\end{equation}
Here, $\tan\beta_1=v_2/v_1$ and $\tan\beta_2=v_3/\sqrt{v_1^2+v_2^2}$. Finally, writing the mass terms for the CP-Odd Higgs in a similar fashion
\begin{equation*}
        V_n^{mass} \supset \begin{pmatrix}
            n_1 & n_2 & n_3
        \end{pmatrix} \frac{\mathcal{M}^2_n}{2} \begin{pmatrix}
            n_1 \\ n_2 \\ n_3 
        \end{pmatrix}, 
\end{equation*}
We can diagonalize this matrix in a manner similar to the charged Higgs sector. The mass eigenstates are given by
\begin{equation}
    \begin{split}
        G_0 & = (c_{\beta 1}c_{\beta 2}) n_1 + (c_{\beta 2}s_{\beta 1}) n_2 + (s_{\beta 2}) n_3, \\
        A_1 & = (-c_{\gamma 1 }s_{\beta 1} + c_{\beta 1}s_{\beta 2}s_{\gamma 1})n_1 + (c_{\beta 1}c_{\gamma 1} + s_{\beta 1}s_{\beta 2}s_{\gamma 1})n_2 + (-c_{\beta 2}s_{\gamma 1})n_3,\,\textrm{and} \\
        A_2 & = (-c_{\beta 1}c_{\gamma 1}s_{\beta 2}-s_{\beta 1}s_{\gamma 1})n_1 + (-c_{\gamma 1}s_{\beta 1}s_{\beta 2}+c_{\beta 1}s_{\gamma 1})n_2 + (c_{\beta 2}c_{\gamma 1}) n_3.
    \end{split}
\end{equation}
The remaining $\lambda_{7,8,...12}$ terms of the potential are traded off in terms of the charged Higgs and CP-Odd Higgs masses,vevs and the mixing angles.
\subsection{Yukawa Sector}
Experimental limits place stringent constraints on the presence of tree-level Flavor Changing Neutral Currents (FCNCs), posing a significant challenge for multi-Higgs models. To address this, we impose Natural Flavor Conservation (NFC), a condition that restricts each Higgs doublet to couple with only one class of fermions. In the framework we adopt for the present work, we employ the Type-Z or democratic Yukawa structure, wherein the masses of the leptons, the down-type quarks, and the up-type quarks are generated by the $\Phi_1$, $\Phi_2$, and $\Phi_3$ doublets respectively.  We can thus write the Yukawa Lagrangian as
\begin{equation}
		\label{eq:yukeq}
		\mathcal{L}_{Yukawa} = -[\bar{L}_L \Phi_1 \mathcal{G}_l l_R+\bar{Q}_L \Phi_2 \mathcal{G}_d d_R + \bar{Q}_L \tilde{\Phi}_3 \mathcal{G}_u u_R  + h.c],
\end{equation}
where the  $\mathcal{G}_f$ are the Yukawa matrices. In terms of the fermion mass matrices they can be written as
\begin{equation*}
		\mathcal{G}_f = \frac{\sqrt{2} \mathcal{M}_f}{v_i}.
\end{equation*}
We work with a $Z_3$ symmetric potential as given in~\autoref{eq:phitrans} - for the Yukawa Lagrangian to remain invariant under the same, the right handed fermion fields transform as 
\begin{equation}
		\label{eq:fermtrans}
		d_R  \rightarrow \omega d_R , \; \; \; \; l_R \rightarrow \omega^2 l_R, \; \; \; \; u_R \rightarrow  u_R.
\end{equation}
\subsection{Alignment-Limit}
The discovery of a SM like Higgs boson necessitates that one of the CP-even scalars in the 3HDM be identified with it. This requirement imposes specific constraints on the parameters of the scalar potential and the mixing angles. In this work, we adopt the Regular Hierarchy configuration \cite{Batra:2025amk}, wherein the lightest CP-even Higgs boson, $H_1$ is identified as the observed SM Higgs. The coupling of the $H_1$ with the SM gauge bosons in the democratic 3HDM is given by
\begin{equation}
g_{HZZ}=\frac{ve^2(c_{\beta 2}c_{\alpha 2}\cos(\alpha_1-\beta_1)+s_{\beta 2}s_{\alpha 2})}{2c_w^2s_w^2},
\end{equation}
and thus the alignment limit condition for the regular hierarchy reads
 \begin{equation}
     	c_{\beta_2}c_{\alpha_2}\cos(\alpha_1-\beta_1) + s_{\beta_2}s_{\alpha_2} = 1.
      \label{eq:al1}
 \end{equation}
Letting $k=\cos(\alpha_1-\beta_1)$, we have 
\begin{itemize}
    \item $k=1\implies \alpha_1 = \beta_1$ and  $\alpha_2 = \beta_2$.
\end{itemize}
The case with $k=-1$ is equivalent to the current scenario up to an inversion, whereas the case with $k\neq 1$ represents a special configuration in which the entire vacuum expectation value (vev) is confined to the third doublet. Therefore, we proceed with our analysis by choosing $k=1$ \cite{Batra:2025amk}.
\subsection{Constraints}
Herein, we list the various theoretical and experimental constraints that are directly relevant to understand the available parameter space of the model. For a more detailed study of how these constraints were imposed in this model, the reader is invited to consult Ref. \cite{Batra:2025amk}.

\begin{enumerate}[label=(\roman*)]
    \item \textbf{Vacuum Stability:} To ensure stability, the scalar potential must be bounded from below in all directions of the field space.\footnote{We remark here that the complete necessary and sufficient conditions for the $Z_3$ symmetric 3HDM are not yet known. The conditions listed in Eq. \ref{eq:stability} should instead be understood as a set of sufficient stability conditions obtained by requiring positivity of the quartic potential along a representative set of field directions in the scalar field space. These conditions are analogous to the well-known copositivity-inspired constraints \cite{Kannike:2012pe}, used in multi-Higgs models when the full analytic BFB solution is unavailable. While they do not exhaust all possible directions in field space, they guarantee stability along all tested directions and therefore ensure that the scalar potential is bounded from below in a physically relevant region of parameter space. Also see Ref \cite{Boto:2021qgu, Faro:2019vcd}.} This translates into a set of bounds on the scalar coupling parameters $\lambda_i$ which are listed below, 
    \begin{equation}
    \label{eq:stability}
    \begin{gathered}
        \lambda_{1} \geq 0, \quad \lambda_{2} \geq 0, \quad \lambda_{3} \geq 0, \\
         \lambda_{4} + 2\sqrt{\lambda_{1}\lambda_{2}} \geq 0, \quad \lambda_{5} + 2\sqrt{\lambda_{1}\lambda_{3}} \geq 0, \quad \lambda_{6} + 2\sqrt{\lambda_{2}\lambda_{3}} \geq 0, \\
         \lambda_{4} + \lambda_{7} + 2\sqrt{\lambda_{1}\lambda_{2}} \geq 0, \quad \lambda_{5} + \lambda_{8} + 2\sqrt{\lambda_{1}\lambda_{3}} \geq 0, \quad \lambda_{6} + \lambda_{9} + 2\sqrt{\lambda_{2}\lambda_{3}} \geq 0.
    \end{gathered}
    \end{equation}
    \item \textbf{Unitarity:} The unitarity constraints are necessary to ensure that the theory remains predictive at high energies. Using the results of \cite{Bento_2022,Boto_2021} for unitarity Constraints in the $Z_3$ symmetric 3HDM, the condition to be imposed on the $21$ eigenvalues $\Lambda_i$ of the relevant scattering matrices is
    \begin{equation*}
        |\Lambda_i| \leq 8\pi.
    \end{equation*} 
    \item \textbf{Perturbativity:} That the model remains perturbatively calculable as a field theory is ensured by imposing on all the scalar couplings the condition $\lambda_i \leq |4\pi|.$

    \item \textbf{EW Precision Tests:} The Peskin-Takeuchi parameters $S$, $T$, and $U$ which capture the effects of new physics on gauge boson propagators impose stringent constraints on all extensions of the SM. We require that these parameters lie within the following allowed ranges:  
    \begin{align*}
        S & = -0.02 \pm 0.10, \\
        T & = 0.03 \pm 0.12,\,\textrm{and} \\
        U & = 0.01 \pm 0.11. 
    \end{align*}
    \item \textbf{BSM Higgs Boson Exclusion:} The exclusion limits on the BSM scalar sector from direct searches conducted at the LHC, LEP, and Tevatron were computed at the $95 \% $ Confidence Level (C.L.) using the \texttt{HiggsBounds-6} framework, interfaced through the \texttt{HiggsTools} package \cite{Bahl_2023} and imposed on the parameter space of the model.
    \item \textbf{SM-Like Higgs Boson Discovery:} The compatibility of the Higgs boson in the model with the discovered SM-like Higgs boson was ascertained using a goodness-of-fit test by \texttt{HiggsSignals-5} via the \texttt{HiggsTools} package.
    \item \textbf{Flavour Physics:} The most stringent bound on the Branching Ratio ($\mathcal{BR}$) of the $B\rightarrow X_s \gamma$ decay using Next-to-Leading Order (NLO) calculations in QCD as discussed in \cite{Borzumati_1998,Akeroyd_2021,Boto_2021} was used to constrain the model.
\end{enumerate}

In what will follow in the subsequent sections, all the benchmark points chosen are those that are allowed after the imposition of these listed theoretical and experimental constraints. We now turn to the question of discovering the heavy Higgs bosons at the ILC.

\section{Phenomenological Analysis}
\label{sec:Pheno}
To begin the study, we first list all possible modes of production and decay of the heavy scalars in this model at the ILC. We then perform a cut-based analysis for each of the considered production modes. Each channel is studied for two benchmark points (BP), with most cases analyzed in the fully hadronic final state. 

\subsection{Relevant Production Modes}
We begin by listing all relevant production channels that can potentially uncover signals of a heavy Higgs in $e^+e^-$ collisions in~\autoref{tab:benchmark}. For each production mode, we have considered two sets of benchmark mass points, denoted as BPX and BPY, where X and Y are numerical labels assigned to clearly distinguish between different benchmark points across all production modes. These benchmarks satisfy all applied constraints and are thus suitable for a detailed analysis. Depending on the production mode, we consider either fully hadronic decays (FHD) or semi-leptonic decays (SLD) - the corresponding cross sections, branching ratios for the relevant signal channels, and the associated background processes will be discussed in the upcoming sections dedicated to the detailed phenomenological analysis.

\begin{table}[h]
    \centering
    \renewcommand{\arraystretch}{1.2}
    \begin{tabular}{c c c c}
        \toprule
        \textbf{Vertex} & \textbf{Production Mode} & \textbf{Final State} & \textbf{Signal Benchmark} \\
        \midrule
        $SS$  & $H_2A_2$ (\ref{SS})   
        & $4b$ & BP1  :~\autoref{tab:BP1BP2-MT} $(FHD)$ \\
        \addlinespace
        $SSS$  & $h_1H_2A_2$ (\ref{SSS})   
        & $6b$ & BP2  :~\autoref{tab:BP3BP4-MT} $(FHD)$ \\
        \midrule
        \multirow{6}{*}{$SSV$} 
        & $H_2H_2Z$ (\ref{h2h2Z:FHD}) 
        & $4b+2j$ & BP3 :~\autoref{tab:BP5BP6-MT} $(FHD)$ \\
        \addlinespace
        & $A_2A_2Z$ (\ref{A2A2Z:FHD}) 
        & $4b+2j$ & BP4  :~\autoref{tab:BP7BP8-MT} $(FHD)$ \\
        \addlinespace
        & $H_2H_2Z$ (\ref{h2h2Z:SLD}) 
        & $4b+2\ell$ & BP5  :~\autoref{tab:BP9BP10-MT} $(SLD)$ \\
        \addlinespace
        & $A_2A_2Z$ (\ref{A2A2Z:SLD}) 
        & $4b+2\ell$ & BP6 :~\autoref{tab:BP11BP12-MT} $(SLD)$ \\
        \addlinespace
        & $H_2H_2^{\pm}W^{\mp}$ (\ref{h2h2+w-}) 
        & $4b+4j$ & BP7  :~\autoref{tab:BP13BP14-MT} $(FHD)$ \\
        \addlinespace
        & $A_2H_2^{\pm}W^{\mp}$ (\ref{A2h2+w-}) 
        & $4b+4j$ & BP8  :~\autoref{tab:BP15BP16-MT} $(FHD)$ \\
        \bottomrule
    \end{tabular}
    \caption{Benchmark scenarios for different vertices and production modes, outlined with shorthand notations and corresponding table references.}
    \label{tab:benchmark}
\end{table}

The 3HDM, as mentioned in~\autoref{sec:Model}, has multiple neutral and charged Higgs bosons. The choice of the specific channels in the table above is dictated by the parameter space that is allowed after all constraints have been imposed, and also demanding that the relevant couplings - and the associated cross-sections and branching ratios - be large enough that one can in principle probe them at colliders. This also explains why we have chosen a varied set of benchmark points.
\subsection{$H_2A_2$ Production: $FHD$ Mode }
\label{SS}
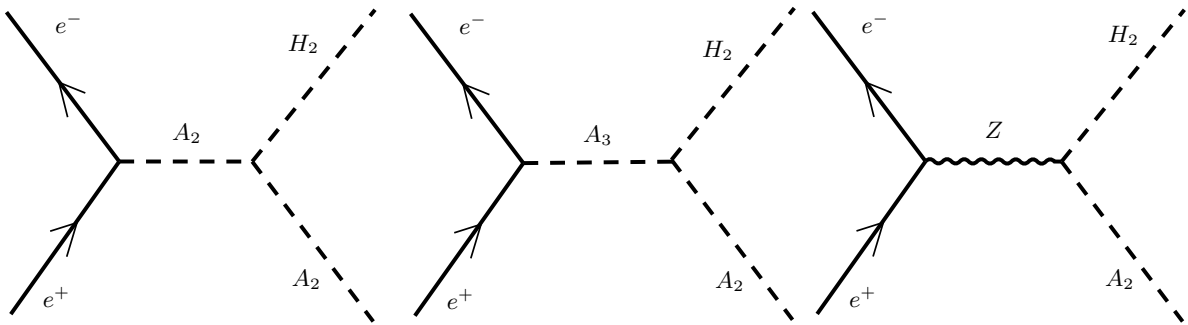
\begin{figure}
    \centering
\tikzset{every picture/.style={line width=0.65pt}} 

\begin{tikzpicture}[x=0.65pt,y=0.65pt,yscale=-1,xscale=1]

\draw [line width=1.5]  [dash pattern={on 5.63pt off 4.5pt}]  (69.62,201.46) -- (147,201.5) ;
\draw [line width=1.5]  [dash pattern={on 5.63pt off 4.5pt}]  (147,201.5) -- (218.46,113.1) ;
\draw [line width=1.5]  [dash pattern={on 5.63pt off 4.5pt}]  (147,201.5) -- (217.44,294.65) ;
\draw [line width=1.5]    (69.62,201.46) -- (4,114.5) ;
\draw [line width=1.5]    (6.51,290.37) -- (69.62,201.46) ;
\draw   (29.41,242.05) -- (45.01,237.1) -- (39.05,256.92) ;
\draw   (39.58,175.65) -- (34.51,155.38) -- (49.87,161.57) ;
\draw [line width=1.5]  [dash pattern={on 5.63pt off 4.5pt}]  (304.77,202.4) -- (397,201.5) ;
\draw [line width=1.5]  [dash pattern={on 5.63pt off 4.5pt}]  (391,200.5) -- (462.46,112.1) ;
\draw [line width=1.5]  [dash pattern={on 5.63pt off 4.5pt}]  (391,200.5) -- (461.44,293.65) ;
\draw [line width=1.5]    (304.77,202.4) -- (239.15,115.43) ;
\draw [line width=1.5]    (241.66,291.31) -- (304.77,202.4) ;
\draw   (264.56,242.99) -- (280.17,238.03) -- (274.21,257.86) ;
\draw   (274.74,176.59) -- (269.67,156.32) -- (285.03,162.51) ;
\draw [line width=1.5]    (538.86,201.33) .. controls (540.53,199.67) and (542.19,199.67) .. (543.85,201.34) .. controls (545.52,203.01) and (547.18,203.01) .. (548.85,201.35) .. controls (550.52,199.69) and (552.18,199.69) .. (553.85,201.36) .. controls (555.52,203.03) and (557.18,203.03) .. (558.85,201.37) .. controls (560.52,199.71) and (562.19,199.72) .. (563.85,201.39) .. controls (565.52,203.06) and (567.18,203.06) .. (568.85,201.4) .. controls (570.52,199.74) and (572.18,199.74) .. (573.85,201.41) .. controls (575.52,203.08) and (577.18,203.08) .. (578.85,201.42) .. controls (580.52,199.76) and (582.18,199.76) .. (583.85,201.43) .. controls (585.52,203.1) and (587.18,203.1) .. (588.85,201.44) .. controls (590.52,199.78) and (592.18,199.78) .. (593.85,201.45) .. controls (595.52,203.12) and (597.18,203.12) .. (598.85,201.46) .. controls (600.52,199.8) and (602.18,199.8) .. (603.85,201.47) .. controls (605.52,203.14) and (607.18,203.14) .. (608.85,201.48) .. controls (610.52,199.82) and (612.18,199.82) .. (613.85,201.49) -- (618,201.5) -- (618,201.5) ;
\draw [line width=1.5]  [dash pattern={on 5.63pt off 4.5pt}]  (618,201.5) -- (689.46,113.1) ;
\draw [line width=1.5]  [dash pattern={on 5.63pt off 4.5pt}]  (618,201.5) -- (688.44,294.65) ;
\draw [line width=1.5]    (538.86,201.33) -- (473.24,114.37) ;
\draw [line width=1.5]    (475.75,290.25) -- (538.86,201.33) ;
\draw   (498.64,241.92) -- (514.25,236.97) -- (508.29,256.79) ;
\draw   (508.82,175.52) -- (503.75,155.26) -- (519.11,161.44) ;

\draw (23.51,272.97) node [anchor=north west][inner sep=0.75pt]    {$e^{+}$};
\draw (30.62,113.09) node [anchor=north west][inner sep=0.75pt]    {$e^{-}$};
\draw (98.76,178.06) node [anchor=north west][inner sep=0.75pt]    {$A_{2}$};
\draw (166.08,125.8) node [anchor=north west][inner sep=0.75pt]    {$H_{2}$};
\draw (168.41,262.96) node [anchor=north west][inner sep=0.75pt]    {$A_{2}$};
\draw (258.66,273.9) node [anchor=north west][inner sep=0.75pt]    {$e^{+}$};
\draw (265.78,114.03) node [anchor=north west][inner sep=0.75pt]    {$e^{-}$};
\draw (337.91,179) node [anchor=north west][inner sep=0.75pt]    {$A_{3}$};
\draw (407.23,129.74) node [anchor=north west][inner sep=0.75pt]    {$H_{2}$};
\draw (414.78,264.05) node [anchor=north west][inner sep=0.75pt]    {$A_{2}$};
\draw (492.74,272.84) node [anchor=north west][inner sep=0.75pt]    {$e^{+}$};
\draw (499.86,112.96) node [anchor=north west][inner sep=0.75pt]    {$e^{-}$};
\draw (571.75,176.77) node [anchor=north west][inner sep=0.75pt]    {$Z$};
\draw (643.32,120.67) node [anchor=north west][inner sep=0.75pt]    {$H_{2}$};
\draw (641.7,262.99) node [anchor=north west][inner sep=0.75pt]    {$A_{2}$};

\end{tikzpicture}
    \caption{A few illustrative Feynman diagrams for the $H_2A_2$ process.}
    \label{fig:H2A2FD}
\end{figure}

For the $H_2 A_2$ production mode, we consider the fully hadronic decay channel which in this case means that both $H_2$ and $A_2$ decay to $b\bar{b}$ resulting in a $4b$ final state. As previously mentioned, we study the efficacy of this channel (like all channels to follow) with a benchmark point (BP), with the corresponding branching ratios (BR) and total cross sections $(\sigma)$ provided in~\autoref{tab:BP1BP2-MT}. In~\autoref{fig:H2A2FD}, we display a few illustrative Feynman diagrams for the production process. For the purpose of phenomenological analysis, we also identify the most relevant background processes—both reducible and irreducible—which are listed in~\autoref{tab:BP1BP2-BG}.

\begin{table}[h]
    \centering
    \renewcommand{\arraystretch}{1.2}
    \begin{tabular}{c c c c c c c c c c c c c}
        \toprule
        BP1:Particles $\rightarrow$ 
        & $h_1$ & $H_2$ & $H_3$ & $A_2$ & $A_3$ & $H_2^{\pm}$ & $H_3^{\pm}$ 
        & Channel 
        & $e^+ e^- \rightarrow H_2A_2$ 
        & $H_2 \rightarrow b \bar{b}$ 
        & $A_2 \rightarrow b \bar{b}$ 
        & Total $\sigma$ (pb)  \\
        \midrule
        Mass (GeV) 
        & 125 & 340.46 & 409.61 & 274.85 & 469.73 & 376.15 & 400.29 
        & $\sigma \times BR$ 
        & 0.00598 & 0.75197 & 0.70063 & 0.0031528 \\
        \midrule
        \bottomrule
    \end{tabular}
    \caption{The table corresponds to the BP1 mass points, BR and total cross section ($\sigma$), for the $H_2A_2$ production mode and its fully hadronic decay.}
    \label{tab:BP1BP2-MT}
\end{table}

\begin{table}[h]
    \centering
    \begin{tabular}{l c c}
        \toprule
        \textbf{Background Process} & \textbf{Final State} & \textbf{Cross Section ($\sigma$) [pb]} \\ 
        \midrule
        $e^+e^- \rightarrow b\bar{b}b\bar{b}$ 
        & $4b$ & 0.006746  \\ 
        $e^+e^- \rightarrow t \bar{t} \rightarrow b\bar{b}jjjj$ 
        & $2b+4j$ & 0.06767   \\ 
        $e^+e^- \rightarrow t \bar{t} h \rightarrow b\bar{b}b\bar{b}jjjj$ 
        & $4b+4j$ & 0.0006778  \\ 
        \bottomrule
    \end{tabular}
    \caption{Relevant Standard Model background processes for \( e^+ e^- \to H_2 A_2 \), with \( H_2 \to b\bar{b} \) and \( A_2 \to b\bar{b} \).}
    \label{tab:BP1BP2-BG}
\end{table}

\begin{figure}[h]
    \centering
    \includegraphics[width=0.49\linewidth]{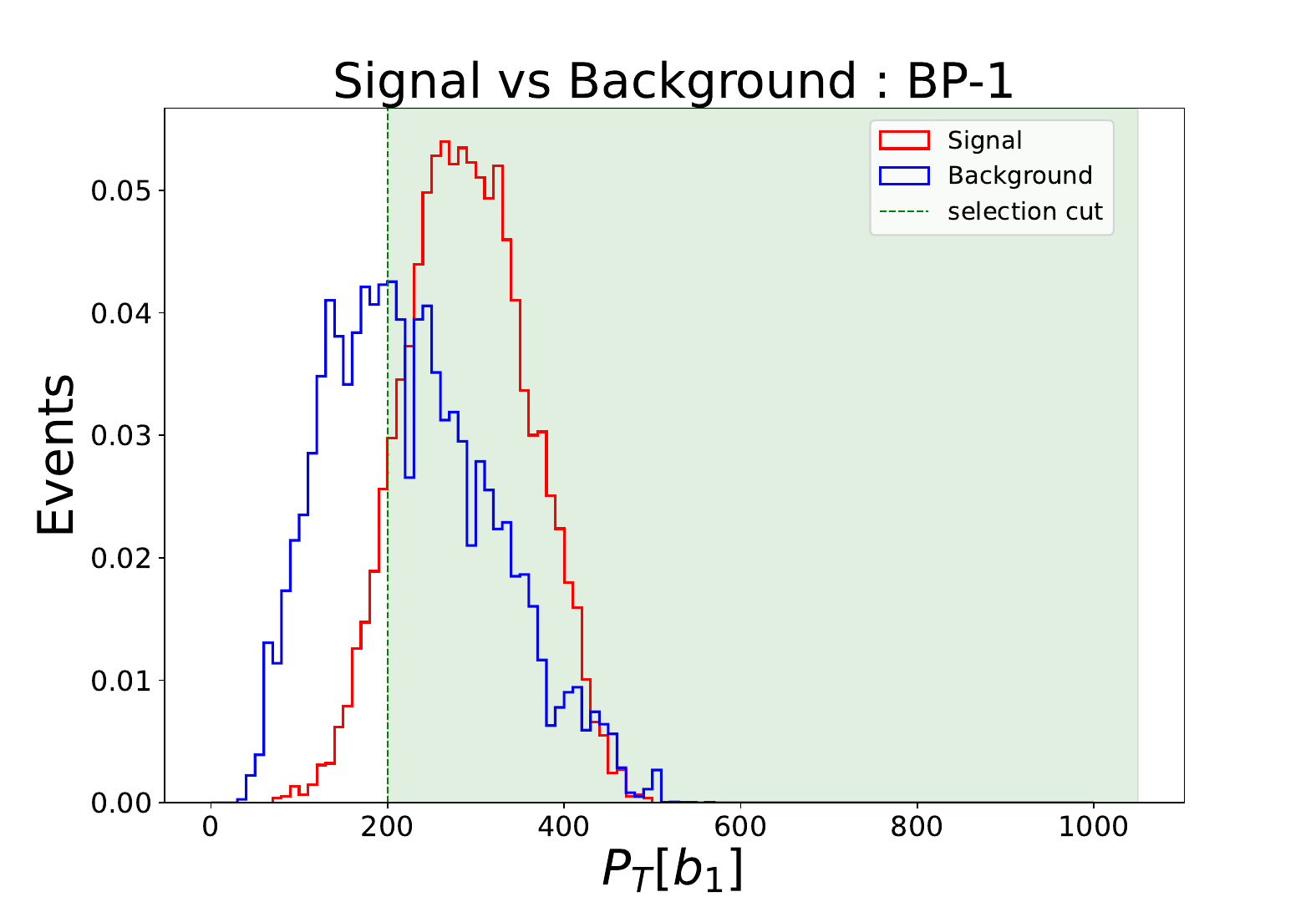}
    \includegraphics[width=0.49\linewidth]{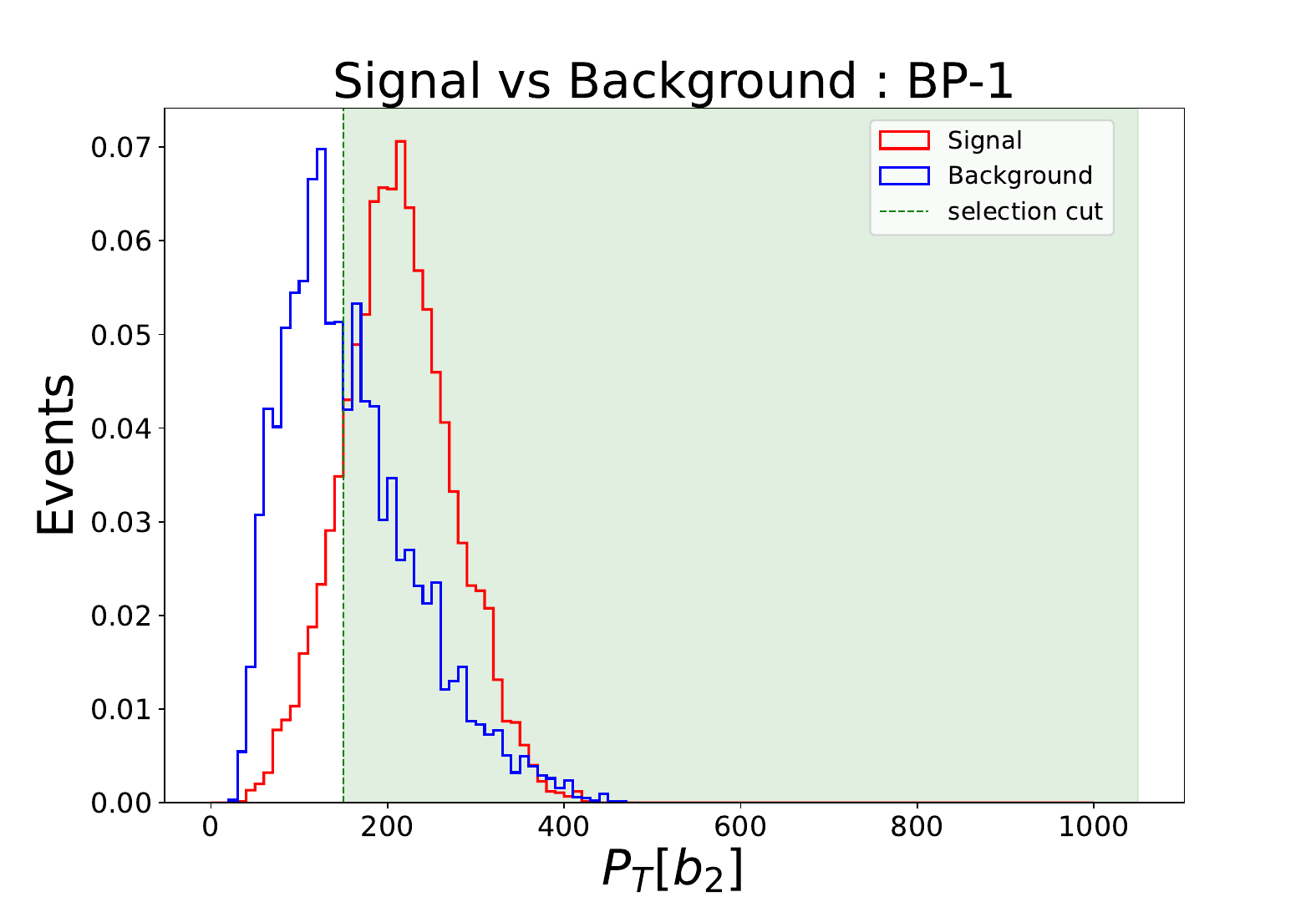}

    \caption{The plots in the upper panel represent the selection cuts applied to BP1. Both sets of plots show the selection cuts on \( p_T(b_1) \) and \( p_T(b_2) \). The shaded regions indicate the selected ranges that are implemented in the cutflow chart.}
    \label{fig:BP1BP2-FIG}
\end{figure}

To isolate the signal in our analysis, we impose a stringent pre-selection cut of $N(b) = 4$\footnote{An alternative strategy could involve a more inclusive requirement such as $N(b) \geq 4$. In realistic detector environments, additional jets may arise due to initial-state radiation (ISR), final-state radiation (FSR), and jet fragmentation. Such effects can result in a higher jet multiplicity than the parton-level expectation. Therefore, using $N(b) \geq 4$ may be beneficial in scenarios where a strict cut significantly reduces signal efficiency. However, in the present case, this issue is not severe, especially when weighed against the benefit of reduced background efficiency, and the exact $N(b) = 4$ selection is sufficient and effective.}, thereby selecting only events that reconstruct exactly four $b$-tagged jets. Following this pre-selection, we apply additional cuts on the transverse momenta of the leading and subleading $b$-jets, $P_T(b_1)$ and $P_T(b_2)$, for both BP1 and BP2 as both these kinematic variables show sufficient promise in isolating signal from background as can be seen from~\autoref{fig:BP1BP2-FIG}.

The total cross sections for the two chosen benchmark points are of the same order, with only a small difference, making it reasonable to anticipate discovery under similar luminosity conditions. Furthermore, upon examining the observables after applying the \( N(b) = 4 \) cut, the nature of their distinguishability remains comparable. Consequently, we have imposed identical cuts on the same observables, \( p_T(b_1) \) and \( p_T(b_2) \).  The corresponding cutflow chart is presented in~\autoref{tab:BP1BP2-CF}, where the significance is reported in the ``S vs B'' column using the formula \( \frac{S}{\sqrt{S + B}} \),\footnote{Throughout the discussion, the significance shown in the cutflow charts is consistently calculated using the formula \( \frac{S}{\sqrt{S + B}} \).} where $S$ and $B$ denote the number of signal and background events respectively. It is found that, at a luminosity of \( 100\,\mathrm{fb}^{-1} \), this channel exhibits a discovery potential of \( 5\sigma \)  based on the implemented selection cuts. However, given that the kinematic properties of the four $b$ quarks are rather similar, it would not be possible to identify which pair belongs to the $H_2$ decay and which ones to the $A_2$, and thus, we do not attempt mass reconstruction. However, even identifying the $4b$ channel and imposing a minimal set of cuts helps us isolate BSM effects quite clearly in this scenario.

\begin{table}[h]
    \centering
    \begin{tabular}{c c c c}
        \toprule
        \textbf{BP1 : Cuts} & \textbf{Signal (S)} & \textbf{Background (B)} & \textbf{S vs B} \\ 
        \midrule
        Initial (no cut) & 315 & 7509 & --- \\ 
        $N(b)=4$ & 58.82 & 85.11 & 4.903 \\ 
        SEL: $P_T(b_1) > 200$ & 53.14 & 48.24 & 5.278 \\ 
        SEL: $P_T(b_2) > 150$ & 47.27 & 35.20 & 5.206 \\ 
        \bottomrule
    \end{tabular}
    \caption{The table represents the cutflow chart for BP1. This is considered at an integrated luminosity of \( 100\,\mathrm{fb}^{-1} \) in the analysis. In this setup, a discovery potential of \( 5\sigma \) is anticipated for the signal.}
    \label{tab:BP1BP2-CF}
\end{table}



\subsection{$h_1H_2A_2$ Production : $FHD$ Mode}
\label{SSS}

This particular channel involving the production of \( h_1 H_2 A_2 \) and its subsequent decay into a fully hadronic mode with a \( 6b \) quark final state offers another promising avenue for exploration. The benchmark point considered here is referred to as BP2 - the corresponding masses, along with information regarding the branching ratios and total cross sections, are provided in~\autoref{tab:BP3BP4-MT}. 

A distinctive feature of this channel is its rare final state consisting of six \( b \)-quarks. The full production and decay chain is given below:
\begin{equation*}
    e^+ e^- \rightarrow h_1 H_2 A_2 \quad ( h_1 \rightarrow b \bar{b}) \quad ( H_2 \rightarrow b \bar{b}) \quad ( A_2 \rightarrow b \bar{b})
    \label{eq2}
\end{equation*}
Due to the unusual final state compared to SM backgrounds, the signal may appear particularly clean, potentially reducing the need for cuts on observable distributions such as the momenta of b-quarks, especially if a strong pre-selection cut has already been applied. Based on the final state feature, a strong requirement of \( N(b) = 6 \) is imposed\footnote{We have implemented stringent preselection cuts here, where the multiplicity requirement exactly matches the partonic-level expectation. This does not significantly affect the signal efficiency, especially when weighed against the resulting reduction in background efficiency; otherwise, a slightly looser cut such as \( N(b) \geq 6 \) would have been considered.
} to effectively isolate signal events. In~\autoref{tab:BP3BP4-BG}, we list both irreducible and reducible backgrounds that are relevant for this channel along with their cross-sections.

\begin{table}[h]
    \centering
    \renewcommand{\arraystretch}{1.2}
    \begin{tabular}{c c c c c c c c c c c c c c}
        \toprule
        BP2:Particle 
        & $h_1$ & $H_2$ & $H_3$ & $A_2$ & $A_3$ & $H_2^{\pm}$ & $H_3^{\pm}$ 
        & Channel 
        & $e^+ e^- \rightarrow h_1H_2A_2$ 
        & $h_1 \rightarrow b \bar{b}$ 
        & $H_2 \rightarrow b \bar{b}$ 
        & $A_2 \rightarrow b \bar{b}$  
        & Total $\sigma$ (pb) \\
        \midrule
        Mass (GeV) 
        & 125 & 255.20 & 543.43 & 228.65 & 366.61 & 347.99 & 452.39 
        & $\sigma \times BR$ 
        & 0.0012179 & 0.54458 & 0.98325 & 0.981178 & 0.00063986 \\
        \bottomrule
    \end{tabular}
    \caption{The upper portion of the table represents the mass points, BR and total cross section corresponding to BP2.}
    \label{tab:BP3BP4-MT}
\end{table}

\begin{table}[h]
    \centering
    \begin{tabular}{l c c}
        \toprule
        \textbf{Background Process} & \textbf{Final State} & \textbf{Cross Section ($\sigma$) [pb]} \\ 
        \midrule
        $e^+e^- \rightarrow b\bar{b}b\bar{b}b\bar{b}$ 
        & $6b$ & $5.886 \times 10^{-6}$  \\ 
        $e^+e^- \rightarrow t \bar{t} \rightarrow b\bar{b}jjjj$ 
        & $2b+4j$ & $0.06767$  \\
        $e^+e^- \rightarrow t \bar{t} j \rightarrow b\bar{b}jjjjj$ 
        & $2b+5j$ & $0.02346$  \\
        $e^+e^- \rightarrow t \bar{t} h \rightarrow b\bar{b}b\bar{b}jjjj$ 
        & $4b+4j$ & $6.778 \times 10^{-4}$  \\ 
        $e^+e^- \rightarrow t \bar{t} Z \rightarrow b\bar{b}b\bar{b}jjjj$ 
        & $4b+4j$ & $2.774 \times 10^{-4}$ \\ 
        \bottomrule
    \end{tabular}
    \caption{The relevant reducible and irreducible SM backgrounds for the $6b$ final state considered in this channel.}
    \label{tab:BP3BP4-BG}
\end{table}

Given the highly distinctive nature of the final state, which is significantly different from typical SM backgrounds, it is reasonable to focus solely on counting the number of signal and background events based on a multiplicity cut of \( N(b) = 6 \). At this stage, with the implementation of strong preselection cuts, it is highly unlikely that any of the SM backgrounds survive, as such a final state is relatively rare in $e^+e^-$ machines. The goal is to determine the range of luminosity required to effectively probe the production mechanism. We find that, for BP2, an integrated luminosity of \( 1000\,\mathrm{fb}^{-1} \) are sufficient to achieve a discovery significance of \( 5\sigma \), based purely on the event count after applying the \( N(b) = 6 \) selection. The corresponding selection cut and its effect on the event yield are summarized in the cutflow chart presented in~\autoref{tab:BP3BP4-CF}.

\begin{table}[h]
    \centering
    \begin{tabular}{c c c c}
        \toprule
        \textbf{BP2 : Cuts} & \textbf{Signal (S)} & \textbf{Background (B)} & \textbf{S vs B} \\ 
        \midrule
        Initial (no cut) & 639 & 92091 & --- \\ 
        SEL: $N(b) = 6$ & 30.71 & 5.0 & 5.14 \\ 
        \bottomrule
    \end{tabular}
    \caption{The upper portion of the table corresponds to the cutflow chart for BP3 at $\mathcal{L}= 1000\,fb^{-1}$, while the lower portion corresponds to BP4  at $\mathcal{L}=1200\,fb^{-1}$.}
    \label{tab:BP3BP4-CF}
\end{table}



\subsection{$H_2H_2Z$ Production : $FHD$ Mode}
\label{h2h2Z:FHD}

We now extend our discussion to production modes that yield final states involving not only $b$-quarks but also additional light jets. In analyzing such final states - comprising various combinations of $b$-jets and light jets - it is essential to account for realistic detector-level effects. Although the parton-level topology predicts a specific number of jets, additional jets may be produced due to QCD radiation processes \cite{Ellis:1996mzs} such as initial state radiation (ISR), final state radiation (FSR), and jet fragmentation. Conversely, some jets may be too soft to be reconstructed or may fall outside the detector acceptance. Additionally, jets originating from hadronic decays of $W$ or $Z$ bosons may merge due to their collimated nature. To ensure robust signal retention while accounting for these effects, we adopt an inclusive jet multiplicity requirement of $N(j) \geq N_{\text{signal}}$, where $N_{\text{signal}}$ corresponds to the number of jets expected from the parton level signal topology. This approach captures events with additional QCD activity while maintaining consistency with the underlying signal hypothesis\footnote{In previous scenarios such as $H_2A_2$ and $H_1H_2A_2$ production, we implemented cuts that matching with the exact parton level expectations. However, in the present case, doing so leads to a significant loss in signal efficiency relative to background rejection. To mitigate this, we adopt a more inclusive strategy that accounts for ISR, FSR, and jet fragmentation effects.}. 

We now proceed with the analysis of the $H_2 H_2 Z$ production channel in the FHD mode, which results in a final state topology of $4b + 2j$. The specific signal process is
\begin{equation*}
    e^+ e^- \to H_2 H_2 Z, \quad (H_2 \to b\bar{b}), \quad (H_2 \to b\bar{b}), \quad (Z \to jj).
    \label{eq5}
\end{equation*}
Although the signal contains exactly four $b$-quarks at the parton level, detector level reconstruction is impacted by $b$-tagging inefficiencies and possible contamination from gluon splitting processes. Accordingly, a requirement of $N(b) \geq 4$ is imposed to retain signal efficiency while suppressing background contributions, and is adopted along with a cut of $N(j) \geq 2$. Following the structure of the previously discussed channels, we consider a benchmark point for the analysis, denoted as BP3, which is consistent with current experimental constraints. The relevant SM backgrounds for this process are listed in~\autoref{tab:bg_H2H2Z-A2A2Z-HD}.\footnote{This same background table is also referenced in the $A_2 A_2 Z$ production case, as both processes in the FHD mode yield the same final state of $4b + 2j$.} Detailed information on the chosen benchmark points—including their masses, branching ratios, and total production cross sections—is presented in~\autoref{tab:BP5BP6-MT}.

\begin{table}[h]
    \centering
    \begin{tabular}{l c c}
        \toprule
        \textbf{Background Process} & \textbf{Final State} & \textbf{Cross Section ($\sigma$) [pb]} \\ 
        \midrule
        $e^+e^- \rightarrow b\bar{b}b\bar{b}jj$ 
        & $4b+2j$ & 0.00141  \\ 
        $e^+e^- \rightarrow W^+W^-b\bar{b}b\bar{b} \rightarrow b\bar{b}b\bar{b}jjjj$ 
        & $4b+4j$ & 0.0004943   \\ 
        $e^+e^- \rightarrow t \bar{t} h \rightarrow b\bar{b}b\bar{b}jjjj$ 
        & $4b+4j$ & 0.0006778  \\ 
        $e^+e^- \rightarrow t \bar{t} Z \rightarrow b\bar{b}b\bar{b}jjjj$ 
        & $4b+4j$ & 0.0002774  \\ 
        $e^+e^- \rightarrow t \bar{t} \rightarrow b\bar{b}jjjj$ 
        & $2b+4j$ & 0.06767  \\
        $e^+e^- \rightarrow t \bar{t} j \rightarrow b\bar{b}jjjjj$ 
        & $2b+5j$ & 0.02346  \\
        $e^+e^- \rightarrow t \bar{t} jj \rightarrow b\bar{b}jjjjjj$ 
        & $2b+6j$ & 0.005821 \\
        $e^+e^- \rightarrow t \bar{t} jjj \rightarrow b\bar{b}jjjjjjj$ 
        & $2b+7j$ & 0.001098  \\
        \bottomrule
    \end{tabular}
    \caption{The relevant SM backgrounds for the fully hadronic decay mode of \( H_2 H_2 Z \) (\autoref{h2h2Z:FHD}) and \( A_2 A_2 Z \) (\autoref{A2A2Z:FHD}) production.}
    \label{tab:bg_H2H2Z-A2A2Z-HD}
\end{table}

\begin{table}[h]
    \centering
    \renewcommand{\arraystretch}{1.2}
    \begin{tabular}{c c c c c c c c c c c c c c}
        \toprule
        BP3:Particle
        & $h_1$ & $H_2$ & $H_3$ & $A_2$ & $A_3$ & $H_2^{\pm}$ & $H_3^{\pm}$
        & Channel
        & $e^+ e^- \rightarrow H_2H_2Z$
        & $H_2 \rightarrow b \bar{b}$
        & $H_2 \rightarrow b \bar{b}$
        & $Z \rightarrow jj$
        & Total $\sigma$ (pb) \\
        \midrule
        Mass (GeV)
        & 125 & 207.83 & 527.73 & 315.17 & 397.45 & 333.29 & 417.01
        & $\sigma \times BR$
        & 0.007175 & 0.815618 & 0.815618 & 0.548 & 0.002616 \\
        \bottomrule
    \end{tabular}
    \caption{The table corresponds to BP3, summarizing the benchmark mass points along with the associated BR and total cross sections.}
    \label{tab:BP5BP6-MT}
\end{table}

\begin{figure}
    \centering
    \includegraphics[width=0.4\linewidth]{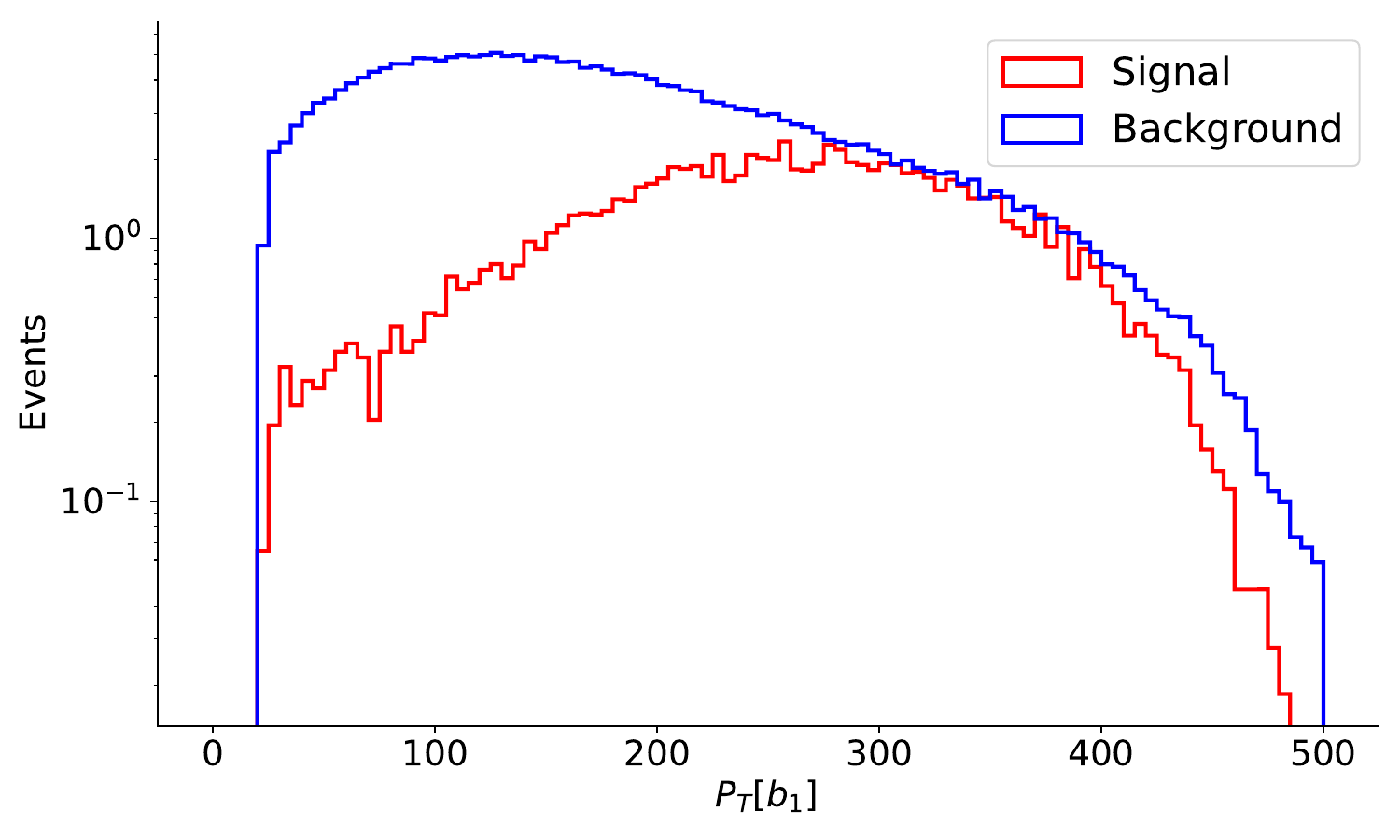}
    \includegraphics[width=0.4\linewidth]{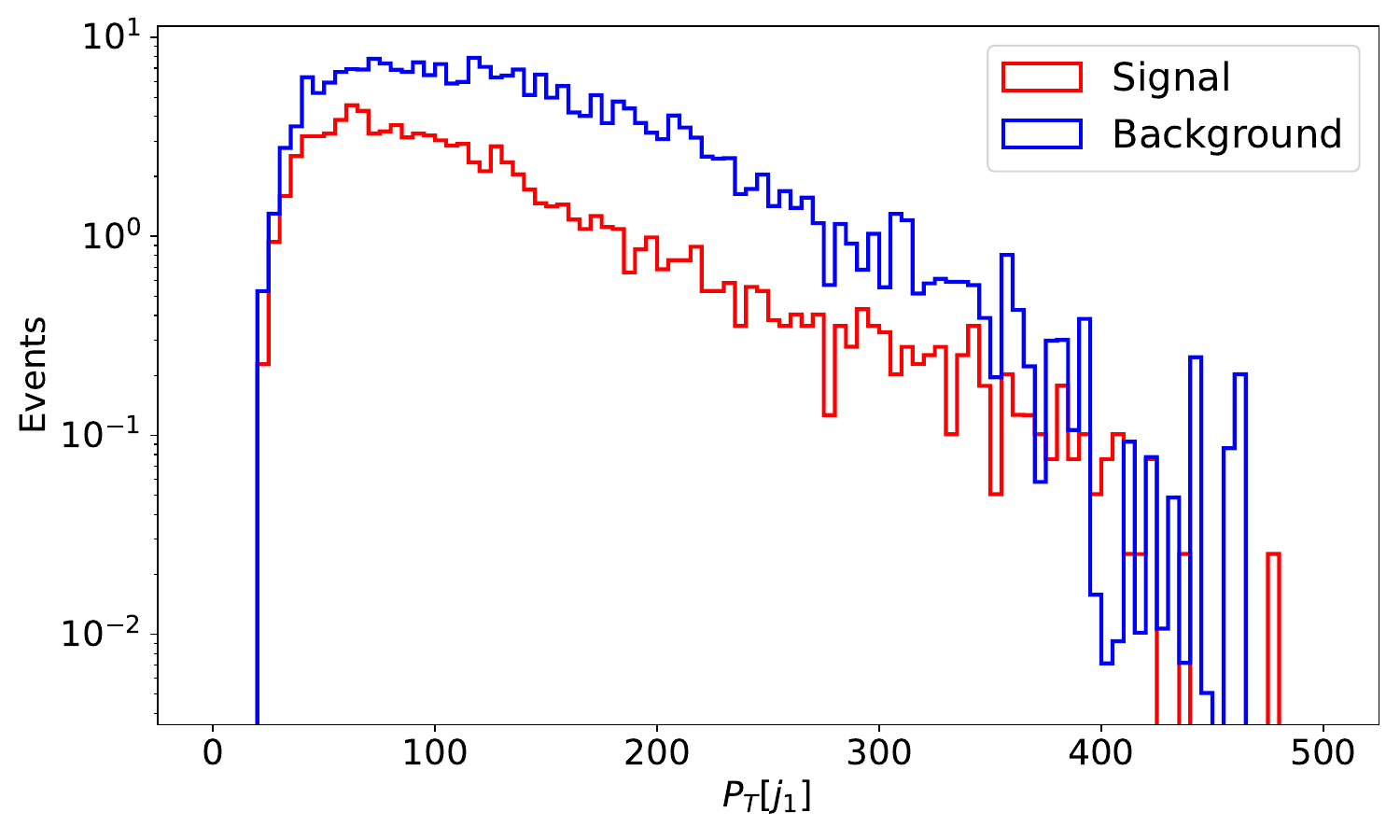}
    \includegraphics[width=0.4\linewidth]{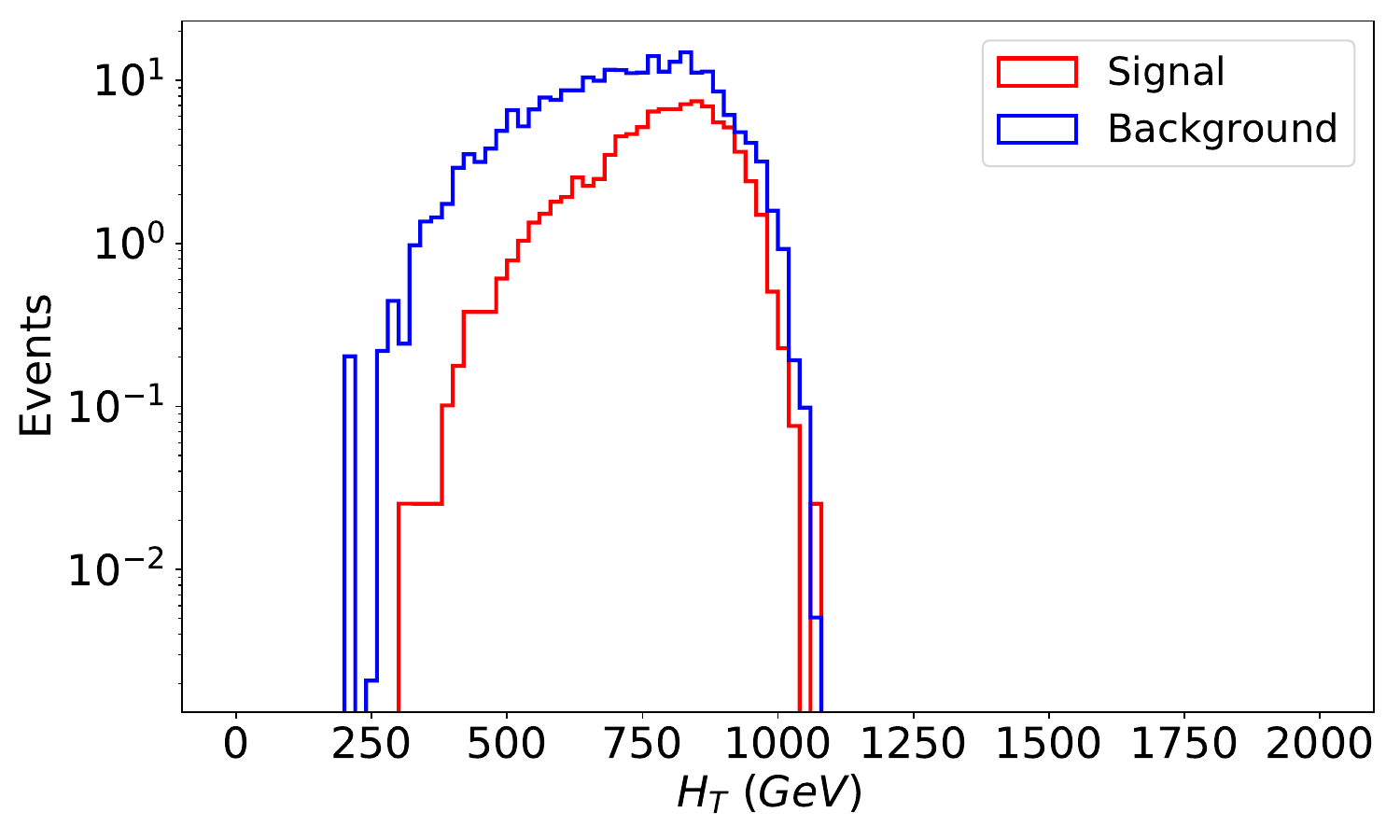}
    \caption{Some sample kinematic distributions for the $H_2H_2Z$ decaying to a fully hadronic final state - as can be seen, the overlap of the signal and background distributions for much of the kinematic range does not permit a useful cut-based strategy in this case.}
    \label{fig:observables}
\end{figure}

The selection cuts \( N(b) \geq 4 \) and \( N(j) \geq 2 \) are employed to preferentially isolate events consistent with the signal topology, thereby suppressing contributions from SM backgrounds. This approach, while simple in implementation, proves effective in enhancing the signal purity of the sample. Nevertheless, an important consequence of these inclusive number-based cuts is the significant overlap observed in the distributions of key kinematic observables between signal and background as can be evidenced from~\autoref{fig:observables}. As a result, no single observable emerges as a strong discriminator suitable for additional shape-based selection. Despite this limitation, the overall event yield remains sufficiently high. At an integrated luminosity of $500\,\mathrm{fb}^{-1}$, the signal reaches the threshold for discovery with a statistical significance of \( 5\sigma \). The cut-based selection criteria data is provided in~\autoref{tab:BP5BP6-CF}. 

\begin{table}[h]
    \centering
    \begin{tabular}{c c c c}
        \toprule
        \textbf{BP3 : Cuts} & \textbf{Signal (S)} & \textbf{Background (B)} & \textbf{S vs B} \\ 
        \midrule
        Initial (no cut) & 1307 & 50454 & --- \\ 
        SEL: \( N(b) \geq 4 \) \& \( N(j) \geq 2 \) & 125.4 & 411.6 & 5.410 \\ 
        \bottomrule
    \end{tabular}
    \caption{The panel presents the cutflow summary for the BP3 signal. Here we assume an integrated luminosity of $500\,\mathrm{fb}^{-1}$.}
    \label{tab:BP5BP6-CF}
\end{table}


\subsection{$A_2A_2Z$ Production : $FHD$ Mode}
\label{A2A2Z:FHD}

The $A_2 A_2 Z$ production channel and its subsequent decay topology are identical to the $H_2 H_2 Z$ case discussed in~\autoref{h2h2Z:FHD}. The corresponding SM backgrounds are listed in~\autoref{tab:bg_H2H2Z-A2A2Z-HD}. The signal benchmarks, labeled as BP7 and BP8, correspond to different mass configurations. Detailed information on their mass values, BR's and total production cross sections is provided in~\autoref{tab:BP7BP8-MT}. The signal process is given by:
\begin{equation*}
    e^+ e^- \to A_2 A_2 Z, \quad (A_2 \to b\bar{b}), \quad (A_2 \to b\bar{b}), \quad (Z \to jj)
    \label{eq6}
\end{equation*}
A similar strategy is adopted for signal isolation in this channel, guided by the final state topology of \( 4b + 2j \). In particular, number-based selection criteria of \( N(b) \geq 4 \) and \( N(j) \geq 2 \) are applied, consistent with the methodology outlined in the previous section. The cutflow summary for benchmark points BP4 is provided in~\autoref{tab:BP7BP8-CF}. Notably, even with only these inclusive number cuts, a favorable balance between signal efficiency and background suppression is achieved, highlighting the effectiveness of this minimal selection strategy.

\begin{table}[h]
    \centering
    \renewcommand{\arraystretch}{1.2}
    \begin{tabular}{c c c c c c c c c c c c c c}
        \toprule
        BP4:Particle
        & $h_1$ & $H_2$ & $H_3$ & $A_2$ & $A_3$ & $H_2^{\pm}$ & $H_3^{\pm}$
        & Channel
        & $e^+ e^- \rightarrow A_2A_2Z$
        & $A_2 \rightarrow b \bar{b}$
        & $A_2 \rightarrow b \bar{b}$
        & $Z \rightarrow jj$
        & Total $\sigma$ (pb) \\
        \midrule
        Mass (GeV)
        & 125 & 354.97 & 552.52 & 210.87 & 391.04 & 369.26 & 404.86
        & $\sigma \times BR$
        & 0.00637189 & 0.8188199 & 0.8188199 & 0.548 & 0.002341 \\
        \bottomrule
    \end{tabular}
    \caption{The table represents the mass points, BR and total cross section information for BP4.}
    \label{tab:BP7BP8-MT}
\end{table}

\begin{table}[h!]
    \centering
    \begin{tabular}{c c c c}
        \toprule
        \textbf{BP4 : Cuts} & \textbf{Signal (S)} & \textbf{Background (B)} & \textbf{S vs B} \\ 
        \midrule
        Initial (no cut) & 936 & 40363 & --- \\
        SEL: $N(b) \geq 4$ \& $N(j) \geq 2$ & 113.42 & 329.3 & 5.391 \\
        \bottomrule
    \end{tabular}
    \caption{The upper panel presents the cutflow summary for the BP4 signal. This case is analyzed at an integrated luminosity of $400\,\mathrm{fb}^{-1}$.}
    \label{tab:BP7BP8-CF}
\end{table}


\subsection{$H_2H_2Z$ Production : $SLD$ Mode}
\label{h2h2Z:SLD}

The $H_2 H_2 Z$ production channel discussed earlier in~\autoref{h2h2Z:FHD} can also be considered with an alternative decay mode, wherein the $Z$ boson decays leptonically. The signal process is thus given by:
\begin{equation*}
    e^+ e^- \to H_2 H_2 Z, \quad (H_2 \to b\bar{b}), \quad (H_2 \to b\bar{b}), \quad (Z \to \ell^+ \ell^-)
    \label{eq7}
\end{equation*}
This leads to a distinct final state of $4b + 2\ell$. One of the consequences could be a drop in the total cross section, as the $Z$ decaying hadronically yields a higher branching ratio, whereas its leptonic decay significantly reduces the total cross section due to a much lower branching ratio. Nevertheless, it must be compared against distinct and appropriate SM backgrounds. To proceed with the analysis, it is necessary to identify Standard Model backgrounds that can mimic this signature. The relevant backgrounds are listed in~\autoref{tab:bg_H2H2Z-A2A2Z-SL}\footnote{The background processes listed apply equally to both $H_2 H_2 Z$ and $A_2 A_2 Z$ production channels in the semi-leptonic decay mode, as they yield identical final states.}. Details of the benchmark point, BP5, including their masses, branching ratios, and cross sections, are provided in~\autoref{tab:BP9BP10-MT}.

\begin{table}[h]
    \centering
    \begin{tabular}{l c c}
        \toprule
        \textbf{Background Process} & \textbf{Final State} & \textbf{Cross Section $\sigma$ (pb)} \\
        \midrule
        $e^+e^- \rightarrow b\bar{b}b\bar{b}l^+l^-$ 
        & $4b+2l$ 
        & $1.893 \times 10^{-5}$ \\

        $e^+e^- \rightarrow W^+W^-b\bar{b}b\bar{b} \rightarrow b\bar{b}b\bar{b}l^+l^- \nu_l \bar{\nu}_l$ 
        & $4b+2l+\slashed{E}_T$ 
        & $5.495 \times 10^{-5}$ \\

        $e^+e^- \rightarrow t\bar{t}h \rightarrow b\bar{b}b\bar{b}l^+l^- \nu_l \bar{\nu}_l$ 
        & $4b+2l+\slashed{E}_T$ 
        & $7.524 \times 10^{-5}$ \\

        $e^+e^- \rightarrow t\bar{t}Z \rightarrow b\bar{b}b\bar{b}l^+l^- \nu_l \bar{\nu}_l$ 
        & $4b+2l+\slashed{E}_T$ 
        & $3.088 \times 10^{-5}$ \\

        $e^+e^- \rightarrow t\bar{t} \rightarrow b\bar{b}l^+l^- \nu_l \bar{\nu}_l$ 
        & $2b+2l+\slashed{E}_T$ 
        & $7.521 \times 10^{-3}$ \\

        $e^+e^- \rightarrow t\bar{t}j \rightarrow b\bar{b}l^+l^- \nu_l \bar{\nu}_l j$ 
        & $2b+2l+\slashed{E}_T+1j$ 
        & $2.605 \times 10^{-3}$ \\

        $e^+e^- \rightarrow t\bar{t}jj \rightarrow b\bar{b}l^+l^- \nu_l \bar{\nu}_l jj$ 
        & $2b+2l+\slashed{E}_T+2j$ 
        & $6.453 \times 10^{-4}$ \\

        $e^+e^- \rightarrow t\bar{t}jjj \rightarrow b\bar{b}l^+l^- \nu_l \bar{\nu}_l jjj$ 
        & $2b+2l+\slashed{E}_T+3j$ 
        & $1.223 \times 10^{-4}$ \\

        $e^+e^- \rightarrow t\bar{t}W^+W^- \rightarrow b\bar{b}b\bar{b}4l\,4\nu_l$ 
        & $4b+4l+\slashed{E}_T$ 
        & $2.913 \times 10^{-7}$ \\

        $e^+e^- \rightarrow t\bar{t}ZZ \rightarrow b\bar{b}b\bar{b}4l\,4\nu_l$ 
        & $4b+4l+\slashed{E}_T$ 
        & $7.712 \times 10^{-9}$ \\

        $e^+e^- \rightarrow t\bar{t}ZZ \rightarrow b\bar{b}b\bar{b}b\bar{b}l^+l^- \nu_l \bar{\nu}_l$ 
        & $6b+2l+\slashed{E}_T$ 
        & $3.365 \times 10^{-8}$ \\
        \bottomrule
    \end{tabular}
    \caption{The relevant SM backgrounds for the semi-leptonic decay mode of $H_2H_2Z$ (\autoref{h2h2Z:SLD}) and $A_2A_2Z$ (\autoref{A2A2Z:SLD}) channels. The demand of at least two leptons necessitates the presence of SM gauge bosons that might also decay to neutrinos, and thus many of these backgrounds have sizable missing energy.}
    \label{tab:bg_H2H2Z-A2A2Z-SL}
\end{table}

\begin{table}[h]
    \centering
    \renewcommand{\arraystretch}{1.2}
    \begin{tabular}{c c c c c c c c c c c c c c}
        \toprule
        BP5:Particle
        & $h_1$ & $H_2$ & $H_3$ & $A_2$ & $A_3$ & $H_2^{\pm}$ & $H_3^{\pm}$
        & Channel
        & $e^+ e^- \rightarrow H_2H_2Z$
        & $H_2 \rightarrow b \bar{b}$
        & $H_2 \rightarrow b \bar{b}$
        & $Z \rightarrow l^+l^-$
        & Total $\sigma$ (pb) \\
        \midrule
        Mass (GeV)
        & 125 & 207.83 & 527.73 & 315.17 & 397.45 & 333.29 & 417.01
        & $\sigma || BR$
        & 0.007175 & 0.815618 & 0.815618 & 0.0673 & 0.0003212 \\
        \bottomrule
    \end{tabular}
    \caption{The table represents the mass points, BR, and total cross-section information for BP5.}
    \label{tab:BP9BP10-MT}
\end{table}

The analysis begins by applying number-based selection criteria, requiring \( N(b) \geq 4 \) and \( N(\ell) \geq 2 \), to facilitate effective reconstruction of the signal. Here, in this case as well, once the required number cuts are implemented, the remaining observable distributions for the signal and backgrounds become highly overlapping, making it difficult to distinguish between them. Also given that the BR for leptonic decays is significantly lower than that of the hadronic mode discussed earlier, a higher integrated luminosity is generally required to achieve comparable signal sensitivity. The cutflow summary for benchmark points \( BP5 \) is presented in~\autoref{tab:BP9BP10-CF}. 

\begin{table}[h]
    \centering
    \renewcommand{\arraystretch}{1.2}
    \begin{tabular}{c c c c}
        \toprule
        \textbf{BP5 : Cuts} & \textbf{Signal (S)} & \textbf{Background (B)} & \textbf{S vs B} \\ 
        \midrule
        Initial (no cut) & 963 & 33221 & --- \\
        SEL: \( N(b) \geq 4 \) $\&$ \( N(\ell) \geq 2 \) & 41.26 & 25.6 & 5.047 \\
        \bottomrule
    \end{tabular}
    \caption{The panel presents the cutflow summary for the BP5 signal. Here we assume an integrated luminosity of $3000\,\mathrm{fb}^{-1}$.}
    \label{tab:BP9BP10-CF}
\end{table}

\subsection{$A_2A_2Z$ Production : $SLD$ Mode}
\label{A2A2Z:SLD}
 
The $A_2 A_2 Z$ production channel followed by a semi-leptonic decay mode is analogous to the process discussed in~\autoref{h2h2Z:SLD}, yielding the same final state. Consequently, the relevant backgrounds remain unchanged and are listed in~\autoref{tab:bg_H2H2Z-A2A2Z-SL}. A benchmark mass point, denoted as BP6, is considered, with its detailed properties provided in~\autoref{tab:BP11BP12-MT}. The signal process is given by:
\begin{equation*}
    e^+ e^- \to A_2 A_2 Z, \quad (A_2 \to b\bar{b}), \quad (A_2 \to b\bar{b}), \quad (Z \to \ell^+ \ell^-)
    \label{eq8}
\end{equation*}
This process yields a final state of \( 4b + 2\ell \). To enable effective signal reconstruction while suppressing Standard Model backgrounds, we impose number-based selection cuts of \( N(b) \geq 4 \) and \( N(\ell) \geq 2 \). The resulting cutflow summary, following the implementation of these selection criteria, is presented in~\autoref{tab:BP11BP12-CF}. 

\begin{table}[h]
    \centering
    \renewcommand{\arraystretch}{1.2}
    \begin{tabular}{c c c c c c c c c c c c c c}
        \toprule
        BP6:Particle
        & $h_1$ & $H_2$ & $H_3$ & $A_2$ & $A_3$ & $H_2^{\pm}$ & $H_3^{\pm}$
        & Channel
        & $e^+ e^- \rightarrow A_2A_2Z$
        & $A_2 \rightarrow b \bar{b}$
        & $A_2 \rightarrow b \bar{b}$
        & $Z \rightarrow l^+l^-$
        & Total $\sigma$ (pb) \\
        \midrule
        Mass (GeV)
        & 125 & 354.97 & 552.52 & 210.87 & 391.04 & 369.26 & 404.86
        & $\sigma \times BR$
        & 0.00637189 & 0.8188199 & 0.8188199 & 0.0673 & 0.0002875 \\
        \bottomrule
    \end{tabular}
    \caption{The table represents the mass points, BR, and total cross-section values for BP6.}
    \label{tab:BP11BP12-MT}
\end{table}

\begin{table}[h]
    \centering
    \begin{tabular}{c c c c}
        \toprule
        \textbf{BP5 : Cuts} & \textbf{Signal (S)} & \textbf{Background (B)} & \textbf{S vs B} \\ 
        \midrule
        Initial (no cut) & 575 & 22147 & --- \\
        SEL: \( N(b) \geq 4 \) $\&$ \( N(\ell) \geq 2 \) & 37.07 & 17.04 & 5.04 \\
        \bottomrule
    \end{tabular}
    \caption{The table presents the cutflow summary for the BP6 signal. Here we assume an integrated luminosity of $2000\,\mathrm{fb}^{-1}$.}
    \label{tab:BP11BP12-CF}
\end{table}

With these findings, we now move on to discussing more non-trivial production modes and their prospects, particularly in probing vertices involving the presence of charged Higgs bosons, as we continue our phenomenological discussions.

\subsection{$H_2H_2^\pm W^\mp$ Production : $FHD$ Mode}
\label{h2h2+w-}

The \( H_2 H_2^{\pm} W^{\mp} \) production mode introduces a charged Higgs boson in the production mode\footnote{This scenario bears a close resemblance to the \( A_2 H_2^{\pm} W^{\mp} \) channel in the FHD mode discussed in~\autoref{A2h2+w-}.}. This particular channel features a more non-trivial and complex final state, with a higher jet multiplicity than those discussed so far. The processe culminate to a final state topology of \( 4b + 4j \). While this signature is relatively rare in SM processes, certain backgrounds can still mimic it due to effects such as mistagging and QCD jet production. We identify and compile a unified set of relevant SM backgrounds, which are listed in~\autoref{tab:bgH2H2+W-_A2H2+W-}.\footnote{This background table is also applicable to the subsequent analysis of the \( A_2 H_2^{\pm} W^{\mp} \) process in the FHD mode.} The signal process for the \( H_2 H_2^{\pm} W^{\mp} \) production channel proceeds through the following decay chain:
\begin{equation*}
    e^+ e^- \to H_2 H_2^{\pm} W^{\mp}, \quad (H_2 \to b\bar{b}), \quad (H_2^{\pm} \to t\bar{b}/\bar{t}b \to W^+ b\bar{b}/W^-b\bar{b} \to j j b \bar{b}), \quad (W^{\mp} \to j j)
    \label{eq3}
\end{equation*}
A representative benchmark point, BP7 is considered in this study. Their mass configurations, BR and total production cross sections are summarized in~\autoref{tab:BP13BP14-MT}. Given the complexity of the \(4b + 4j\) final state and the presence of fully hadronic decays, our selection strategy aims to keep as much signal as possible while reducing the background as much as we can. To this end, we apply the number-based selection cuts \( N(b) \geq 4 \) and \( N(j) \geq 4 \). These cuts are well-aligned with the parton-level expectations of the signal process and are effective in suppressing SM backgrounds, particularly those lacking sufficient $b$-quark content. Although the hadronic nature of the decay and the low signal cross section present challenges, we explore the discovery potential of this channel under a high-luminosity scenario. Our analysis indicates that, with an integrated luminosity of \(2000~\text{fb}^{-1}\), the signal can be clearly discovered with a significance of \(5\sigma\). The cutflow chart corresponding to this analysis, based on cut-based selections, is shown in~\autoref{tab:BP13BP14-CF}.

\begin{table}[t]
\centering
\begin{tabular}{l c c}
\hline\hline
\textbf{Background Process} & \textbf{Final State} & \textbf{Cross Section $\sigma$ (pb)} \\
\hline
$e^{+}e^{-} \rightarrow ZZ\,b\bar{b}b\bar{b} \rightarrow jjjj\,b\bar{b}b\bar{b}$ 
& $4b + 4j$ & $2.089 \times 10^{-7}$ \\

$e^{+}e^{-} \rightarrow W^{+}W^{-} b\bar{b}b\bar{b} \rightarrow jjjj\,b\bar{b}b\bar{b}$ 
& $4b + 4j$ & $4.943 \times 10^{-4}$ \\

$e^{+}e^{-} \rightarrow t\bar{t}h \rightarrow b\bar{b}b\bar{b} jjjj$ 
& $4b + 4j$ & $6.778 \times 10^{-4}$ \\

$e^{+}e^{-} \rightarrow t\bar{t}Z \rightarrow b\bar{b}b\bar{b} jjjj$ 
& $4b + 4j$ & $2.774 \times 10^{-4}$ \\

$e^{+}e^{-} \rightarrow t\bar{t} \rightarrow b\bar{b} jjjj$ 
& $2b + 4j$ & $6.767 \times 10^{-2}$ \\

$e^{+}e^{-} \rightarrow t\bar{t}j \rightarrow b\bar{b} jjjjj$ 
& $2b + 5j$ & $2.346 \times 10^{-2}$ \\

$e^{+}e^{-} \rightarrow t\bar{t}jj \rightarrow b\bar{b} jjjjjj$ 
& $2b + 6j$ & $5.821 \times 10^{-3}$ \\

$e^{+}e^{-} \rightarrow t\bar{t}jjj \rightarrow b\bar{b} jjjjjjj$ 
& $2b + 7j$ & $1.098 \times 10^{-3}$ \\
\hline
\end{tabular}
\caption{Relevant Standard Model backgrounds for the \( H_2 H_2^{\pm} W^{\mp} \) (\autoref{h2h2+w-}) and \( A_2 H_2^{\pm} W^{\mp} \) (\autoref{A2h2+w-}) production channels in the fully hadronic decay ($FHD$) mode, featuring a final state topology of \( 4b + 4j \).}
\label{tab:bgH2H2+W-_A2H2+W-}
\end{table}

\begin{table}[h]
    \centering
    \tiny
    \renewcommand{\arraystretch}{1.2}
    \begin{tabular}{c c c c c c c c c c c c c c c}
        \toprule
        BP7:Particle
        & $h_1$ & $H_2$ & $H_3$ & $A_2$ & $A_3$ & $H_2^{\pm}$ & $H_3^{\pm}$
        & Channel
        & $e^+ e^- \rightarrow H_2 H_2^{\pm} W^{\mp}$
        & $H_2 \rightarrow b \bar{b}$
        & $H_2^{\pm} \rightarrow t \bar{b}/\bar{t}b$
        & $t/\bar{t}\rightarrow W^+b/W^-\bar{b}$
        & $W^{\pm}/W^{\mp} \rightarrow jj$
        & Total $\sigma$ (pb) \\
        \midrule
        Mass (GeV)
        & 125 & 224.16 & 359.96 & 377.98 & 474.54 & 326.93 & 500.67
        & $\sigma \times BR$
        & 0.01095 & 0.8232 & 0.1575 & 1.0 & 0.6 & 0.001022 \\
        \bottomrule
    \end{tabular}
    \caption{The table provides information on \textbf{$H_2 H_2^{\pm} W^{\mp}$} production and its subsequent decay into a \textbf{$4b + 4j$ }final state for BP7.}
    \label{tab:BP13BP14-MT}
\end{table}


\subsection{$A_2H_2^\pm W^\mp$ Production : $FHD$ Mode}
\label{A2h2+w-}

The $A_2 H_2^{\pm} W^{\mp}$ production mode shares its final state topology with the previously discussed $H_2 H_2^{\pm} W^{\mp}$ process, both culminating in a fully hadronic signature of $4b + 4j$. This overlap in decay structure naturally suggests a unified treatment of Standard Model backgrounds, particularly those capable of mimicking this rare final state via mistagging. The common background processes relevant to both channels are listed in~\autoref{tab:bgH2H2+W-_A2H2+W-}. The full production and decay chain under consideration is:
\begin{equation*}
    e^+ e^- \to A_2 H_2^{\pm} W^{\mp}, \quad (A_2 \to b\bar{b}), \quad (H_2^{\pm} \to t\bar{b}/\bar{t}b \to W^+ b\bar{b}/W^-b\bar{b} \to j j b \bar{b}), \quad (W^{\mp} \to j j)
    \label{eq4}
\end{equation*}
We select a benchmark point—denoted as BP8—that satisfy existing phenomenological constraints while offering representative signal kinematics. Their detailed properties, including masses, decay branching fractions, and production cross sections, are provided in~\autoref{tab:BP15BP16-MT}. To extract the signal from the multijet-dominated environment, we impose a set of number-based selection cuts consistent with the parton-level topology: $N(b) \geq 4$ and $N(j) \geq 4$. This inclusive approach accounts for realistic detector effects, such as initial and final-state radiation and jet fragmentation, while maintaining strong discriminatory power against backgrounds lacking sufficient $b$-jet multiplicity. Despite the inherently suppressed cross sections and the challenging hadronic environment, we investigate whether this channel can yield meaningful statistical significance at a high-luminosity collider setup. Our results, summarized in the cutflow~\autoref{tab:BP15BP16-CF}, indicate that with an integrated luminosity of $2000~\text{fb}^{-1}$, the signal can be probed with a discovery-level sensitivity of $5\sigma$.

\begin{table}[h]
    \centering
    \tiny
    \renewcommand{\arraystretch}{1.2}
    \begin{tabular}{c c c c c c c c c c c c c c c}
        \toprule
        BP8:Particle
        & $h_1$ & $H_2$ & $H_3$ & $A_2$ & $A_3$ & $H_2^{\pm}$ & $H_3^{\pm}$
        & Channel
        & $e^+ e^- \rightarrow A_2 H_2^{\pm} W^{\mp}$
        & $A_2 \rightarrow b \bar{b}$
        & $H_2^{\pm} \rightarrow t \bar{b}/\bar{t}b$
        & $t/\bar{t}\rightarrow W^+b/W^-\bar{b}$
        & $W^{\pm}/W^{\mp} \rightarrow jj$
        & Total $\sigma$ (pb) \\
        \midrule
        Mass (GeV)
        & 125 & 365.69 & 370.25 & 222.74 & 366.90 & 309.51 & 437.50
        & $\sigma \times BR$
        & 0.01111 & 0.96079 & 0.28799 & 1.0 & 0.6 & 0.002214 \\
        \bottomrule
    \end{tabular}
    \caption{The table represents the mass points, branching ratios, and total cross-section values for BP8 for the $A_2 H_2^{\pm} W^{\mp} \rightarrow 4b + 4j$ final state.}
    \label{tab:BP15BP16-MT}
\end{table}

\begin{table}[h]
    \centering
    \renewcommand{\arraystretch}{1.2}
    \begin{tabular}{c c c c}
        \toprule
        \textbf{BP7 : Cuts} & \textbf{Signal (S)} & \textbf{Background (B)} & \textbf{S vs B} \\ 
        \midrule
        Initial (no cut) & 2043 & 198997 & -- \\
        SEL: $N(b) \geq 4$ $\&$ $N(j)\geq4$ & 104.0 & 302.8 & 5.15 \\
        \bottomrule
    \end{tabular}
    \caption{Cut-flow results for benchmark points BP7. The signal and background events are analyzed at an integrated luminosity of $2000~\text{fb}^{-1}$.}
    \label{tab:BP13BP14-CF}
\end{table}

\begin{table}[h]
    \centering
    \renewcommand{\arraystretch}{1.2}
    \begin{tabular}{c c c c}
        \toprule
        \textbf{BP8 : Cuts} & \textbf{Signal (S)} & \textbf{Background (B)} & \textbf{S vs B} \\ 
        \midrule
        Initial (no cut) & 4428 & 198997 & -- \\
        SEL: $N(b) \geq 4$ $\&$ $N(j)\geq4$ & 110.24 & 302.8 & 5.43 \\
        \bottomrule
    \end{tabular}
    \caption{Cut-flow results for benchmark points BP8 at an integrated luminosity of $2000~\text{fb}^{-1}$.}
    \label{tab:BP15BP16-CF}
\end{table}


With all these channel surveys carried out, we now move on to the Discussion and Conclusion section~\autoref{sec:Conclusions}, where we highlight the relevant discussions and outline future outlooks that could be considered in experiments such as the ILC.

\section{Discussion and Conclusions}
\label{sec:Conclusions}

Various present and future collider experiments offer distinct capabilities to probe new physics, each with their own unique strengths and experimental setups. The International Linear Collider, in particular, provides a clean experimental environment and energy tunability, which make it especially promising for precision studies of extended Higgs sectors~\cite{Adolphsen:2013kya,ILCInternationalDevelopmentTeam:2022izu,ILC:2013jhg}. Motivated by these advantages, we have investigated the potential of the ILC operating at a center-of-mass energy of $\sqrt{s} = 1000$~GeV to probe physics beyond the SM through an extension of the scalar sector. In this study, we focus on the 3HDM, which provides scope for a rich phenomenology due to its enlarged scalar spectrum and expanded parameter space. The designed framework in the work respects theoretical consistency and complies with current experimental constraints. Our goal was to identify promising search strategies sensitive to interaction vertices characteristic of new physics, and to explore benchmark scenarios where viable production cross sections could be obtained, potentially leading to observable signals.

We performed a detailed analysis of several production channels at the ILC, including $e^+e^- \to H_2 A_2$, $e^+e^- \to H_2 H_2 Z$, $e^+e^- \to A_2 A_2 Z$, $e^+e^- \to H_2 H_2^\pm W^\mp$, $e^+e^- \to A_2 H_2^\pm W^\mp$, and $e^+e^- \to H_1 H_2 A_2$, which yield final states containing multiple jets, leptons, and b-quarks, making them ideal channels to investigate the extended Higgs sector. For each channel, we selected representative benchmark points and conducted detailed phenomenological analysis. It was observed that, with the exception of the $e^+e^- \to H_2 A_2$ channel, the application of additional kinematic variables such as $p_T(j_1)$ or $p_T(b_1)$ beyond the required object multiplicity cuts offers limited improvement in signal sensitivity. This is due to significant overlap between signal and background distributions after basic selection, making further discrimination through such observables less effective.

This study presents a phenomenological survey of several channels through which new physics interaction vertices can be probed. In all the processes considered, at least two new particles contribute to the final state. For example, in $e^+e^- \to H_2 A_2$, both $H_2$ and $A_2$ decay into $b$-quark pairs with overlapping kinematics due to their similar masses, making it difficult to reconstruct them individually from $b$-jet pairings. Nonetheless, an excess of events can still reveal the presence of such interactions. Similar challenges arise in all other channels. While addressing these combinatorial issues is beyond the scope of this work, future studies may explore dedicated strategies for improving signal reconstruction in such scenarios~\cite{Badea:2022dzb,Kim:2021wrr}. However, the channels considered in our study do exhibit promising discovery potential at various luminosity thresholds. With the application of optimized selection cuts and transverse momentum requirements on the leading and subleading $b$-jets, the $H_2A_2$ channel achieves a $5\sigma$ discovery significance at an integrated luminosity of approximately $100~\text{fb}^{-1}$ for benchmark points BP1. For the $h_1H_2A_2$ channel, discovery is attainable at $1000~\text{fb}^{-1}$ for BP2. In the $H_2H_2A_2$ channel, the benchmark point BP3 require $500~\text{fb}^{-1}$, while BP5 need around $3000~\text{fb}^{-1}$ for discovery. The $A_2H_2A_2$ channel reaches $5\sigma$ at $400~\text{fb}^{-1}$ for BP4, and $2000~\text{fb}^{-1}$ for BP6. Finally, channels involving a charged Higgs boson, $H_2H_2^{\pm}W^{\mp}$ and $A_2H_2^{\pm}W^{\mp}$, demand an integrated luminosity close to $2000~\text{fb}^{-1}$ to achieve $5\sigma$ discovery significance for BP7 and BP8. These results collectively demonstrate that multiple new physics channels within this framework can reach discovery sensitivity at the considered collider setup.

\begin{figure}
    \centering
    \includegraphics[width=0.75\linewidth]{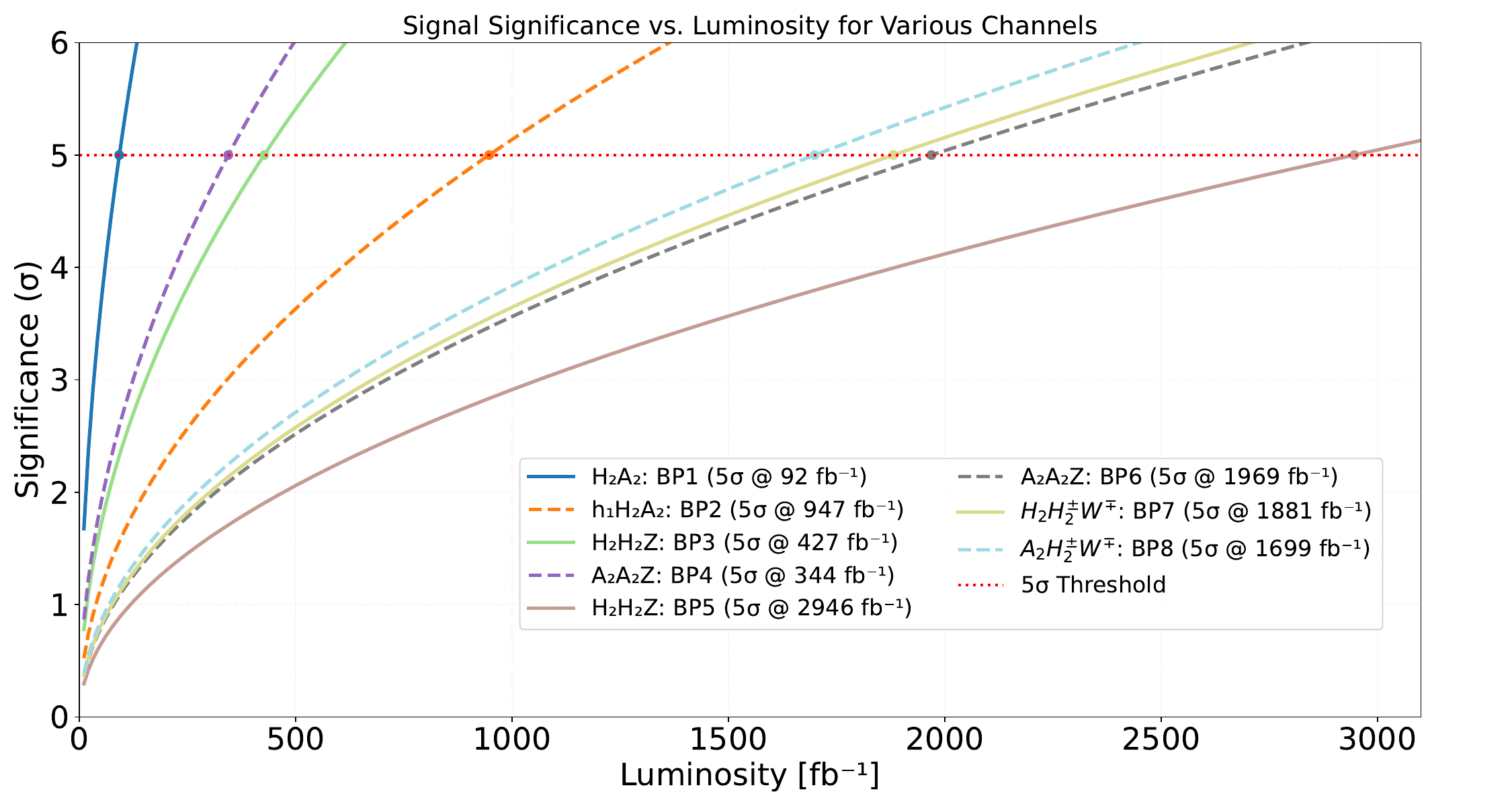}
    \caption{Signal significance vs luminosity for all considered channels.}
    \label{fig:results}
\end{figure}

Overall, this study highlights promising directions for probing extended Higgs sectors at the ILC and providing the information of channel-wise analysis. We have compiled all the production channels considered in this work along with their respective decay modes, as summarized in~\autoref{tab:benchmark}. The discovery prospects of each production mode have been evaluated under different luminosity scenarios, as shown through detailed cutflow analyses\footnote{Most of the cutflow steps involve object multiplicity requirements designed to match the parton-level expectations.}. The resulting signal significances for various channels, compiled across varying luminosities, are presented as the main outcome of this phenomenological survey in~\autoref{fig:results}.

\section{Acknowledgements}

A.K. acknowledges the support from the Director's Fellowship at IIT Gandhinagar. G.B.K.\ acknowledges the initial financial support received under the Prime Minister's Research Fellowship (Grant No.\ PMRF-192002-1802), awarded by the Ministry of Education, Government of India. We thank the anonymous referee for a careful reading of the manuscript and for several helpful comments and suggestions.

\bibliography{reference}

\begin{thebibliography}{41}%
\makeatletter
\providecommand \@ifxundefined [1]{%
 \@ifx{#1\undefined}
}%
\providecommand \@ifnum [1]{%
 \ifnum #1\expandafter \@firstoftwo
 \else \expandafter \@secondoftwo
 \fi
}%
\providecommand \@ifx [1]{%
 \ifx #1\expandafter \@firstoftwo
 \else \expandafter \@secondoftwo
 \fi
}%
\providecommand \natexlab [1]{#1}%
\providecommand \enquote  [1]{``#1''}%
\providecommand \bibnamefont  [1]{#1}%
\providecommand \bibfnamefont [1]{#1}%
\providecommand \citenamefont [1]{#1}%
\providecommand \href@noop [0]{\@secondoftwo}%
\providecommand \href [0]{\begingroup \@sanitize@url \@href}%
\providecommand \@href[1]{\@@startlink{#1}\@@href}%
\providecommand \@@href[1]{\endgroup#1\@@endlink}%
\providecommand \@sanitize@url [0]{\catcode `\\12\catcode `\$12\catcode
  `\&12\catcode `\#12\catcode `\^12\catcode `\_12\catcode `\%12\relax}%
\providecommand \@@startlink[1]{}%
\providecommand \@@endlink[0]{}%
\providecommand \url  [0]{\begingroup\@sanitize@url \@url }%
\providecommand \@url [1]{\endgroup\@href {#1}{\urlprefix }}%
\providecommand \urlprefix  [0]{URL }%
\providecommand \Eprint [0]{\href }%
\providecommand \doibase [0]{https://doi.org/}%
\providecommand \selectlanguage [0]{\@gobble}%
\providecommand \bibinfo  [0]{\@secondoftwo}%
\providecommand \bibfield  [0]{\@secondoftwo}%
\providecommand \translation [1]{[#1]}%
\providecommand \BibitemOpen [0]{}%
\providecommand \bibitemStop [0]{}%
\providecommand \bibitemNoStop [0]{.\EOS\space}%
\providecommand \EOS [0]{\spacefactor3000\relax}%
\providecommand \BibitemShut  [1]{\csname bibitem#1\endcsname}%
\let\auto@bib@innerbib\@empty
\bibitem [{\citenamefont {Aad}\ \emph {et~al.}(2012)\citenamefont {Aad} \emph
  {et~al.}}]{ATLAS:2012yve}%
  \BibitemOpen
  \bibfield  {author} {\bibinfo {author} {\bibfnamefont {G.}~\bibnamefont
  {Aad}} \emph {et~al.} (\bibinfo {collaboration} {ATLAS}),\ }\bibfield
  {title} {\bibinfo {title} {{Observation of a new particle in the search for
  the Standard Model Higgs boson with the ATLAS detector at the LHC}},\ }\href
  {https://doi.org/10.1016/j.physletb.2012.08.020} {\bibfield  {journal}
  {\bibinfo  {journal} {Phys. Lett. B}\ }\textbf {\bibinfo {volume} {716}},\
  \bibinfo {pages} {1} (\bibinfo {year} {2012})},\ \Eprint
  {https://arxiv.org/abs/1207.7214} {arXiv:1207.7214 [hep-ex]} \BibitemShut
  {NoStop}%
\bibitem [{\citenamefont {Chatrchyan}\ \emph {et~al.}(2012)\citenamefont
  {Chatrchyan} \emph {et~al.}}]{CMS:2012qbp}%
  \BibitemOpen
  \bibfield  {author} {\bibinfo {author} {\bibfnamefont {S.}~\bibnamefont
  {Chatrchyan}} \emph {et~al.} (\bibinfo {collaboration} {CMS}),\ }\bibfield
  {title} {\bibinfo {title} {{Observation of a New Boson at a Mass of 125 GeV
  with the CMS Experiment at the LHC}},\ }\href
  {https://doi.org/10.1016/j.physletb.2012.08.021} {\bibfield  {journal}
  {\bibinfo  {journal} {Phys. Lett. B}\ }\textbf {\bibinfo {volume} {716}},\
  \bibinfo {pages} {30} (\bibinfo {year} {2012})},\ \Eprint
  {https://arxiv.org/abs/1207.7235} {arXiv:1207.7235 [hep-ex]} \BibitemShut
  {NoStop}%
\bibitem [{\citenamefont {Barger}\ \emph {et~al.}(2009)\citenamefont {Barger},
  \citenamefont {Langacker}, \citenamefont {McCaskey}, \citenamefont
  {Ramsey-Musolf},\ and\ \citenamefont {Shaughnessy}}]{Barger:2008jx}%
  \BibitemOpen
  \bibfield  {author} {\bibinfo {author} {\bibfnamefont {V.}~\bibnamefont
  {Barger}}, \bibinfo {author} {\bibfnamefont {P.}~\bibnamefont {Langacker}},
  \bibinfo {author} {\bibfnamefont {M.}~\bibnamefont {McCaskey}}, \bibinfo
  {author} {\bibfnamefont {M.}~\bibnamefont {Ramsey-Musolf}},\ and\ \bibinfo
  {author} {\bibfnamefont {G.}~\bibnamefont {Shaughnessy}},\ }\bibfield
  {title} {\bibinfo {title} {{Complex Singlet Extension of the Standard
  Model}},\ }\href {https://doi.org/10.1103/PhysRevD.79.015018} {\bibfield
  {journal} {\bibinfo  {journal} {Phys. Rev. D}\ }\textbf {\bibinfo {volume}
  {79}},\ \bibinfo {pages} {015018} (\bibinfo {year} {2009})},\ \Eprint
  {https://arxiv.org/abs/0811.0393} {arXiv:0811.0393 [hep-ph]} \BibitemShut
  {NoStop}%
\bibitem [{\citenamefont {Tavartkiladze}(2022)}]{Tavartkiladze:2022pzf}%
  \BibitemOpen
  \bibfield  {author} {\bibinfo {author} {\bibfnamefont {Z.}~\bibnamefont
  {Tavartkiladze}},\ }\bibfield  {title} {\bibinfo {title} {{SM extension with
  a gauged flavor U(1)F symmetry}},\ }\href
  {https://doi.org/10.1103/PhysRevD.106.115002} {\bibfield  {journal} {\bibinfo
   {journal} {Phys. Rev. D}\ }\textbf {\bibinfo {volume} {106}},\ \bibinfo
  {pages} {115002} (\bibinfo {year} {2022})},\ \Eprint
  {https://arxiv.org/abs/2209.14404} {arXiv:2209.14404 [hep-ph]} \BibitemShut
  {NoStop}%
\bibitem [{\citenamefont {Branco}\ \emph {et~al.}(2012)\citenamefont {Branco},
  \citenamefont {Ferreira}, \citenamefont {Lavoura}, \citenamefont {Rebelo},
  \citenamefont {Sher},\ and\ \citenamefont {Silva}}]{Branco:2011iw}%
  \BibitemOpen
  \bibfield  {author} {\bibinfo {author} {\bibfnamefont {G.~C.}\ \bibnamefont
  {Branco}}, \bibinfo {author} {\bibfnamefont {P.~M.}\ \bibnamefont
  {Ferreira}}, \bibinfo {author} {\bibfnamefont {L.}~\bibnamefont {Lavoura}},
  \bibinfo {author} {\bibfnamefont {M.~N.}\ \bibnamefont {Rebelo}}, \bibinfo
  {author} {\bibfnamefont {M.}~\bibnamefont {Sher}},\ and\ \bibinfo {author}
  {\bibfnamefont {J.~P.}\ \bibnamefont {Silva}},\ }\bibfield  {title} {\bibinfo
  {title} {{Theory and phenomenology of two-Higgs-doublet models}},\ }\href
  {https://doi.org/10.1016/j.physrep.2012.02.002} {\bibfield  {journal}
  {\bibinfo  {journal} {Phys. Rept.}\ }\textbf {\bibinfo {volume} {516}},\
  \bibinfo {pages} {1} (\bibinfo {year} {2012})},\ \Eprint
  {https://arxiv.org/abs/1106.0034} {arXiv:1106.0034 [hep-ph]} \BibitemShut
  {NoStop}%
\bibitem [{\citenamefont {Bento}\ \emph {et~al.}(2017)\citenamefont {Bento},
  \citenamefont {Haber}, \citenamefont {Rom\~ao},\ and\ \citenamefont
  {Silva}}]{Bento:2017eti}%
  \BibitemOpen
  \bibfield  {author} {\bibinfo {author} {\bibfnamefont {M.~P.}\ \bibnamefont
  {Bento}}, \bibinfo {author} {\bibfnamefont {H.~E.}\ \bibnamefont {Haber}},
  \bibinfo {author} {\bibfnamefont {J.~C.}\ \bibnamefont {Rom\~ao}},\ and\
  \bibinfo {author} {\bibfnamefont {J.~P.}\ \bibnamefont {Silva}},\ }\bibfield
  {title} {\bibinfo {title} {{Multi-Higgs doublet models: physical
  parametrization, sum rules and unitarity bounds}},\ }\href
  {https://doi.org/10.1007/JHEP11(2017)095} {\bibfield  {journal} {\bibinfo
  {journal} {JHEP}\ }\textbf {\bibinfo {volume} {11}},\ \bibinfo {pages}
  {095}},\ \Eprint {https://arxiv.org/abs/1708.09408} {arXiv:1708.09408
  [hep-ph]} \BibitemShut {NoStop}%
\bibitem [{\citenamefont {Bento}\ \emph {et~al.}(2018)\citenamefont {Bento},
  \citenamefont {Haber}, \citenamefont {Rom\~ao},\ and\ \citenamefont
  {Silva}}]{Bento:2018fmy}%
  \BibitemOpen
  \bibfield  {author} {\bibinfo {author} {\bibfnamefont {M.~P.}\ \bibnamefont
  {Bento}}, \bibinfo {author} {\bibfnamefont {H.~E.}\ \bibnamefont {Haber}},
  \bibinfo {author} {\bibfnamefont {J.~C.}\ \bibnamefont {Rom\~ao}},\ and\
  \bibinfo {author} {\bibfnamefont {J.~P.}\ \bibnamefont {Silva}},\ }\bibfield
  {title} {\bibinfo {title} {{Multi-Higgs doublet models: the Higgs-fermion
  couplings and their sum rules}},\ }\href
  {https://doi.org/10.1007/JHEP10(2018)143} {\bibfield  {journal} {\bibinfo
  {journal} {JHEP}\ }\textbf {\bibinfo {volume} {10}},\ \bibinfo {pages}
  {143}},\ \Eprint {https://arxiv.org/abs/1808.07123} {arXiv:1808.07123
  [hep-ph]} \BibitemShut {NoStop}%
\bibitem [{\citenamefont {Batra}\ \emph {et~al.}(2025)\citenamefont {Batra},
  \citenamefont {Coleppa}, \citenamefont {Khanna}, \citenamefont {Rai},\ and\
  \citenamefont {Sarkar}}]{Batra:2025amk}%
  \BibitemOpen
  \bibfield  {author} {\bibinfo {author} {\bibfnamefont {N.}~\bibnamefont
  {Batra}}, \bibinfo {author} {\bibfnamefont {B.}~\bibnamefont {Coleppa}},
  \bibinfo {author} {\bibfnamefont {A.}~\bibnamefont {Khanna}}, \bibinfo
  {author} {\bibfnamefont {S.~K.}\ \bibnamefont {Rai}},\ and\ \bibinfo {author}
  {\bibfnamefont {A.}~\bibnamefont {Sarkar}},\ }\bibfield  {title} {\bibinfo
  {title} {{Constraining the 3HDM Parameter Space}},\ }\href@noop {} {\
  (\bibinfo {year} {2025})},\ \Eprint {https://arxiv.org/abs/2504.07489}
  {arXiv:2504.07489 [hep-ph]} \BibitemShut {NoStop}%
\bibitem [{\citenamefont {Keus}\ \emph {et~al.}(2014)\citenamefont {Keus},
  \citenamefont {King},\ and\ \citenamefont {Moretti}}]{Keus:2013hya}%
  \BibitemOpen
  \bibfield  {author} {\bibinfo {author} {\bibfnamefont {V.}~\bibnamefont
  {Keus}}, \bibinfo {author} {\bibfnamefont {S.~F.}\ \bibnamefont {King}},\
  and\ \bibinfo {author} {\bibfnamefont {S.}~\bibnamefont {Moretti}},\
  }\bibfield  {title} {\bibinfo {title} {{Three-Higgs-doublet models:
  symmetries, potentials and Higgs boson masses}},\ }\href
  {https://doi.org/10.1007/JHEP01(2014)052} {\bibfield  {journal} {\bibinfo
  {journal} {JHEP}\ }\textbf {\bibinfo {volume} {01}},\ \bibinfo {pages}
  {052}},\ \Eprint {https://arxiv.org/abs/1310.8253} {arXiv:1310.8253 [hep-ph]}
  \BibitemShut {NoStop}%
\bibitem [{\citenamefont {Das}\ and\ \citenamefont {Saha}(2019)}]{Das:2019yad}%
  \BibitemOpen
  \bibfield  {author} {\bibinfo {author} {\bibfnamefont {D.}~\bibnamefont
  {Das}}\ and\ \bibinfo {author} {\bibfnamefont {I.}~\bibnamefont {Saha}},\
  }\bibfield  {title} {\bibinfo {title} {{Alignment limit in three
  Higgs-doublet models}},\ }\href {https://doi.org/10.1103/PhysRevD.100.035021}
  {\bibfield  {journal} {\bibinfo  {journal} {Phys. Rev. D}\ }\textbf {\bibinfo
  {volume} {100}},\ \bibinfo {pages} {035021} (\bibinfo {year} {2019})},\
  \Eprint {https://arxiv.org/abs/1904.03970} {arXiv:1904.03970 [hep-ph]}
  \BibitemShut {NoStop}%
\bibitem [{\citenamefont {Boto}\ \emph
  {et~al.}(2023{\natexlab{a}})\citenamefont {Boto}, \citenamefont {Das},
  \citenamefont {Lourenco}, \citenamefont {Romao},\ and\ \citenamefont
  {Silva}}]{Boto:2023nyi}%
  \BibitemOpen
  \bibfield  {author} {\bibinfo {author} {\bibfnamefont {R.}~\bibnamefont
  {Boto}}, \bibinfo {author} {\bibfnamefont {D.}~\bibnamefont {Das}}, \bibinfo
  {author} {\bibfnamefont {L.}~\bibnamefont {Lourenco}}, \bibinfo {author}
  {\bibfnamefont {J.~C.}\ \bibnamefont {Romao}},\ and\ \bibinfo {author}
  {\bibfnamefont {J.~P.}\ \bibnamefont {Silva}},\ }\bibfield  {title} {\bibinfo
  {title} {{Fingerprinting the type-Z three-Higgs-doublet models}},\ }\href
  {https://doi.org/10.1103/PhysRevD.108.015020} {\bibfield  {journal} {\bibinfo
   {journal} {Phys. Rev. D}\ }\textbf {\bibinfo {volume} {108}},\ \bibinfo
  {pages} {015020} (\bibinfo {year} {2023}{\natexlab{a}})},\ \Eprint
  {https://arxiv.org/abs/2304.13494} {arXiv:2304.13494 [hep-ph]} \BibitemShut
  {NoStop}%
\bibitem [{\citenamefont {Weinberg}(1976)}]{PhysRevLett.37.657}%
  \BibitemOpen
  \bibfield  {author} {\bibinfo {author} {\bibfnamefont {S.}~\bibnamefont
  {Weinberg}},\ }\bibfield  {title} {\bibinfo {title} {Gauge theory of
  $\mathrm{CP}$ nonconservation},\ }\href
  {https://doi.org/10.1103/PhysRevLett.37.657} {\bibfield  {journal} {\bibinfo
  {journal} {Phys. Rev. Lett.}\ }\textbf {\bibinfo {volume} {37}},\ \bibinfo
  {pages} {657} (\bibinfo {year} {1976})}\BibitemShut {NoStop}%
\bibitem [{\citenamefont {Ivanov}(2017)}]{Ivanov:2017dad}%
  \BibitemOpen
  \bibfield  {author} {\bibinfo {author} {\bibfnamefont {I.~P.}\ \bibnamefont
  {Ivanov}},\ }\bibfield  {title} {\bibinfo {title} {{Building and testing
  models with extended Higgs sectors}},\ }\href
  {https://doi.org/10.1016/j.ppnp.2017.03.001} {\bibfield  {journal} {\bibinfo
  {journal} {Prog. Part. Nucl. Phys.}\ }\textbf {\bibinfo {volume} {95}},\
  \bibinfo {pages} {160} (\bibinfo {year} {2017})},\ \Eprint
  {https://arxiv.org/abs/1702.03776} {arXiv:1702.03776 [hep-ph]} \BibitemShut
  {NoStop}%
\bibitem [{\citenamefont {Cordero}\ \emph {et~al.}(2018)\citenamefont
  {Cordero}, \citenamefont {Hernandez-Sanchez}, \citenamefont {Keus},
  \citenamefont {King}, \citenamefont {Moretti}, \citenamefont {Rojas},\ and\
  \citenamefont {Sokolowska}}]{Cordero:2017owj}%
  \BibitemOpen
  \bibfield  {author} {\bibinfo {author} {\bibfnamefont {A.}~\bibnamefont
  {Cordero}}, \bibinfo {author} {\bibfnamefont {J.}~\bibnamefont
  {Hernandez-Sanchez}}, \bibinfo {author} {\bibfnamefont {V.}~\bibnamefont
  {Keus}}, \bibinfo {author} {\bibfnamefont {S.~F.}\ \bibnamefont {King}},
  \bibinfo {author} {\bibfnamefont {S.}~\bibnamefont {Moretti}}, \bibinfo
  {author} {\bibfnamefont {D.}~\bibnamefont {Rojas}},\ and\ \bibinfo {author}
  {\bibfnamefont {D.}~\bibnamefont {Sokolowska}},\ }\bibfield  {title}
  {\bibinfo {title} {{Dark Matter Signals at the LHC from a 3HDM}},\ }\href
  {https://doi.org/10.1007/JHEP05(2018)030} {\bibfield  {journal} {\bibinfo
  {journal} {JHEP}\ }\textbf {\bibinfo {volume} {05}},\ \bibinfo {pages}
  {030}},\ \Eprint {https://arxiv.org/abs/1712.09598} {arXiv:1712.09598
  [hep-ph]} \BibitemShut {NoStop}%
\bibitem [{\citenamefont {Dey}\ \emph {et~al.}(2024)\citenamefont {Dey},
  \citenamefont {Keus}, \citenamefont {Moretti},\ and\ \citenamefont
  {Shepherd-Themistocleous}}]{Dey:2023exa}%
  \BibitemOpen
  \bibfield  {author} {\bibinfo {author} {\bibfnamefont {A.}~\bibnamefont
  {Dey}}, \bibinfo {author} {\bibfnamefont {V.}~\bibnamefont {Keus}}, \bibinfo
  {author} {\bibfnamefont {S.}~\bibnamefont {Moretti}},\ and\ \bibinfo {author}
  {\bibfnamefont {C.}~\bibnamefont {Shepherd-Themistocleous}},\ }\bibfield
  {title} {\bibinfo {title} {{A smoking gun signature of the 3HDM}},\ }\href
  {https://doi.org/10.1007/JHEP07(2024)038} {\bibfield  {journal} {\bibinfo
  {journal} {JHEP}\ }\textbf {\bibinfo {volume} {07}},\ \bibinfo {pages}
  {038}},\ \Eprint {https://arxiv.org/abs/2310.06593} {arXiv:2310.06593
  [hep-ph]} \BibitemShut {NoStop}%
\bibitem [{\citenamefont {Ivanov}\ and\ \citenamefont
  {Vdovin}(2013)}]{Ivanov:2012fp}%
  \BibitemOpen
  \bibfield  {author} {\bibinfo {author} {\bibfnamefont {I.~P.}\ \bibnamefont
  {Ivanov}}\ and\ \bibinfo {author} {\bibfnamefont {E.}~\bibnamefont
  {Vdovin}},\ }\bibfield  {title} {\bibinfo {title} {{Classification of finite
  reparametrization symmetry groups in the three-Higgs-doublet model}},\ }\href
  {https://doi.org/10.1140/epjc/s10052-013-2309-x} {\bibfield  {journal}
  {\bibinfo  {journal} {Eur. Phys. J. C}\ }\textbf {\bibinfo {volume} {73}},\
  \bibinfo {pages} {2309} (\bibinfo {year} {2013})},\ \Eprint
  {https://arxiv.org/abs/1210.6553} {arXiv:1210.6553 [hep-ph]} \BibitemShut
  {NoStop}%
\bibitem [{\citenamefont {Ivanov}\ and\ \citenamefont
  {Vdovin}(2012)}]{Ivanov:2012ry}%
  \BibitemOpen
  \bibfield  {author} {\bibinfo {author} {\bibfnamefont {I.~P.}\ \bibnamefont
  {Ivanov}}\ and\ \bibinfo {author} {\bibfnamefont {E.}~\bibnamefont
  {Vdovin}},\ }\bibfield  {title} {\bibinfo {title} {{Discrete symmetries in
  the three-Higgs-doublet model}},\ }\href
  {https://doi.org/10.1103/PhysRevD.86.095030} {\bibfield  {journal} {\bibinfo
  {journal} {Phys. Rev. D}\ }\textbf {\bibinfo {volume} {86}},\ \bibinfo
  {pages} {095030} (\bibinfo {year} {2012})},\ \Eprint
  {https://arxiv.org/abs/1206.7108} {arXiv:1206.7108 [hep-ph]} \BibitemShut
  {NoStop}%
\bibitem [{\citenamefont {Morrissey}\ and\ \citenamefont
  {Ramsey-Musolf}(2012)}]{Morrissey:2012db}%
  \BibitemOpen
  \bibfield  {author} {\bibinfo {author} {\bibfnamefont {D.~E.}\ \bibnamefont
  {Morrissey}}\ and\ \bibinfo {author} {\bibfnamefont {M.~J.}\ \bibnamefont
  {Ramsey-Musolf}},\ }\bibfield  {title} {\bibinfo {title} {{Electroweak
  baryogenesis}},\ }\href {https://doi.org/10.1088/1367-2630/14/12/125003}
  {\bibfield  {journal} {\bibinfo  {journal} {New J. Phys.}\ }\textbf {\bibinfo
  {volume} {14}},\ \bibinfo {pages} {125003} (\bibinfo {year} {2012})},\
  \Eprint {https://arxiv.org/abs/1206.2942} {arXiv:1206.2942 [hep-ph]}
  \BibitemShut {NoStop}%
\bibitem [{\citenamefont {Cline}(2006)}]{Cline:2006ts}%
  \BibitemOpen
  \bibfield  {author} {\bibinfo {author} {\bibfnamefont {J.~M.}\ \bibnamefont
  {Cline}},\ }\bibfield  {title} {\bibinfo {title} {{Baryogenesis}},\ }in\
  \href@noop {} {\emph {\bibinfo {booktitle} {{Les Houches Summer School -
  Session 86: Particle Physics and Cosmology: The Fabric of Spacetime}}}}\
  (\bibinfo {year} {2006})\ \Eprint {https://arxiv.org/abs/hep-ph/0609145}
  {arXiv:hep-ph/0609145} \BibitemShut {NoStop}%
\bibitem [{\citenamefont {Barbieri}\ \emph {et~al.}(2006)\citenamefont
  {Barbieri}, \citenamefont {Hall},\ and\ \citenamefont
  {Rychkov}}]{Barbieri:2006dq}%
  \BibitemOpen
  \bibfield  {author} {\bibinfo {author} {\bibfnamefont {R.}~\bibnamefont
  {Barbieri}}, \bibinfo {author} {\bibfnamefont {L.~J.}\ \bibnamefont {Hall}},\
  and\ \bibinfo {author} {\bibfnamefont {V.~S.}\ \bibnamefont {Rychkov}},\
  }\bibfield  {title} {\bibinfo {title} {{Improved naturalness with a heavy
  Higgs: An Alternative road to LHC physics}},\ }\href
  {https://doi.org/10.1103/PhysRevD.74.015007} {\bibfield  {journal} {\bibinfo
  {journal} {Phys. Rev. D}\ }\textbf {\bibinfo {volume} {74}},\ \bibinfo
  {pages} {015007} (\bibinfo {year} {2006})},\ \Eprint
  {https://arxiv.org/abs/hep-ph/0603188} {arXiv:hep-ph/0603188} \BibitemShut
  {NoStop}%
\bibitem [{\citenamefont {Abelleira~Fernandez}\ \emph
  {et~al.}(2012)\citenamefont {Abelleira~Fernandez} \emph
  {et~al.}}]{LHeCStudyGroup:2012zhm}%
  \BibitemOpen
  \bibfield  {author} {\bibinfo {author} {\bibfnamefont {J.~L.}\ \bibnamefont
  {Abelleira~Fernandez}} \emph {et~al.} (\bibinfo {collaboration} {LHeC Study
  Group}),\ }\bibfield  {title} {\bibinfo {title} {{A Large Hadron Electron
  Collider at CERN: Report on the Physics and Design Concepts for Machine and
  Detector}},\ }\href {https://doi.org/10.1088/0954-3899/39/7/075001}
  {\bibfield  {journal} {\bibinfo  {journal} {J. Phys. G}\ }\textbf {\bibinfo
  {volume} {39}},\ \bibinfo {pages} {075001} (\bibinfo {year} {2012})},\
  \Eprint {https://arxiv.org/abs/1206.2913} {arXiv:1206.2913 [physics.acc-ph]}
  \BibitemShut {NoStop}%
\bibitem [{\citenamefont {Ogreid}\ \emph {et~al.}(2017)\citenamefont {Ogreid},
  \citenamefont {Osland},\ and\ \citenamefont {Rebelo}}]{Ogreid:2017alh}%
  \BibitemOpen
  \bibfield  {author} {\bibinfo {author} {\bibfnamefont {O.~M.}\ \bibnamefont
  {Ogreid}}, \bibinfo {author} {\bibfnamefont {P.}~\bibnamefont {Osland}},\
  and\ \bibinfo {author} {\bibfnamefont {M.~N.}\ \bibnamefont {Rebelo}},\
  }\bibfield  {title} {\bibinfo {title} {{A Simple Method to detect spontaneous
  CP Violation in multi-Higgs models}},\ }\href
  {https://doi.org/10.1007/JHEP08(2017)005} {\bibfield  {journal} {\bibinfo
  {journal} {JHEP}\ }\textbf {\bibinfo {volume} {08}},\ \bibinfo {pages}
  {005}},\ \Eprint {https://arxiv.org/abs/1701.04768} {arXiv:1701.04768
  [hep-ph]} \BibitemShut {NoStop}%
\bibitem [{\citenamefont {Adolphsen}\ \emph {et~al.}(2022)\citenamefont
  {Adolphsen} \emph {et~al.}}]{Adolphsen:2022ibf}%
  \BibitemOpen
  \bibfield  {author} {\bibinfo {author} {\bibfnamefont {C.}~\bibnamefont
  {Adolphsen}} \emph {et~al.},\ }\bibfield  {title} {\bibinfo {title}
  {{European Strategy for Particle Physics -- Accelerator R{\&}D Roadmap}},\
  }\href {https://doi.org/10.23731/CYRM-2022-001} {\bibfield  {journal}
  {\bibinfo  {journal} {CERN Yellow Rep. Monogr.}\ }\textbf {\bibinfo {volume}
  {1}},\ \bibinfo {pages} {1} (\bibinfo {year} {2022})},\ \Eprint
  {https://arxiv.org/abs/2201.07895} {arXiv:2201.07895 [physics.acc-ph]}
  \BibitemShut {NoStop}%
\bibitem [{\citenamefont {Bambade}\ \emph {et~al.}(2019)\citenamefont {Bambade}
  \emph {et~al.}}]{Bambade:2019fyw}%
  \BibitemOpen
  \bibfield  {author} {\bibinfo {author} {\bibfnamefont {P.}~\bibnamefont
  {Bambade}} \emph {et~al.},\ }\bibfield  {title} {\bibinfo {title} {{The
  International Linear Collider: A Global Project}},\ }\href@noop {} {\
  (\bibinfo {year} {2019})},\ \Eprint {https://arxiv.org/abs/1903.01629}
  {arXiv:1903.01629 [hep-ex]} \BibitemShut {NoStop}%
\bibitem [{\citenamefont {Barklow}\ \emph {et~al.}(2015)\citenamefont
  {Barklow}, \citenamefont {Brau}, \citenamefont {Fujii}, \citenamefont {Gao},
  \citenamefont {List}, \citenamefont {Walker},\ and\ \citenamefont
  {Yokoya}}]{Barklow:2015tja}%
  \BibitemOpen
  \bibfield  {author} {\bibinfo {author} {\bibfnamefont {T.}~\bibnamefont
  {Barklow}}, \bibinfo {author} {\bibfnamefont {J.}~\bibnamefont {Brau}},
  \bibinfo {author} {\bibfnamefont {K.}~\bibnamefont {Fujii}}, \bibinfo
  {author} {\bibfnamefont {J.}~\bibnamefont {Gao}}, \bibinfo {author}
  {\bibfnamefont {J.}~\bibnamefont {List}}, \bibinfo {author} {\bibfnamefont
  {N.}~\bibnamefont {Walker}},\ and\ \bibinfo {author} {\bibfnamefont
  {K.}~\bibnamefont {Yokoya}},\ }\bibfield  {title} {\bibinfo {title} {{ILC
  Operating Scenarios}},\ }\href@noop {} {\  (\bibinfo {year} {2015})},\
  \Eprint {https://arxiv.org/abs/1506.07830} {arXiv:1506.07830 [hep-ex]}
  \BibitemShut {NoStop}%
\bibitem [{Ado(2013)}]{Adolphsen:2013kya}%
  \BibitemOpen
  \bibfield  {title} {\bibinfo {title} {{The International Linear Collider
  Technical Design Report - Volume 3.II: Accelerator Baseline Design}},\
  }\href@noop {} {\  (\bibinfo {year} {2013})},\ \Eprint
  {https://arxiv.org/abs/1306.6328} {arXiv:1306.6328 [physics.acc-ph]}
  \BibitemShut {NoStop}%
\bibitem [{\citenamefont {Boto}\ \emph
  {et~al.}(2023{\natexlab{b}})\citenamefont {Boto}, \citenamefont {Das},
  \citenamefont {Lourenco}, \citenamefont {Rom\~ao},\ and\ \citenamefont
  {Silva}}]{PhysRevD.108.015020}%
  \BibitemOpen
  \bibfield  {author} {\bibinfo {author} {\bibfnamefont {R.}~\bibnamefont
  {Boto}}, \bibinfo {author} {\bibfnamefont {D.}~\bibnamefont {Das}}, \bibinfo
  {author} {\bibfnamefont {L.}~\bibnamefont {Lourenco}}, \bibinfo {author}
  {\bibfnamefont {J.~C.}\ \bibnamefont {Rom\~ao}},\ and\ \bibinfo {author}
  {\bibfnamefont {J.~P.}\ \bibnamefont {Silva}},\ }\bibfield  {title} {\bibinfo
  {title} {Fingerprinting the type-z three-higgs-doublet models},\ }\href
  {https://doi.org/10.1103/PhysRevD.108.015020} {\bibfield  {journal} {\bibinfo
   {journal} {Phys. Rev. D}\ }\textbf {\bibinfo {volume} {108}},\ \bibinfo
  {pages} {015020} (\bibinfo {year} {2023}{\natexlab{b}})}\BibitemShut
  {NoStop}%
\bibitem [{\citenamefont {Boto}\ \emph
  {et~al.}(2021{\natexlab{a}})\citenamefont {Boto}, \citenamefont {Rom{\~a}o},\
  and\ \citenamefont {Silva}}]{Boto:2021qgu}%
  \BibitemOpen
  \bibfield  {author} {\bibinfo {author} {\bibfnamefont {R.}~\bibnamefont
  {Boto}}, \bibinfo {author} {\bibfnamefont {J.~C.}\ \bibnamefont
  {Rom{\~a}o}},\ and\ \bibinfo {author} {\bibfnamefont {J.~P.}\ \bibnamefont
  {Silva}},\ }\bibfield  {title} {\bibinfo {title} {{Current bounds on the
  type-Z Z3 three-Higgs-doublet model}},\ }\href
  {https://doi.org/10.1103/PhysRevD.104.095006} {\bibfield  {journal} {\bibinfo
   {journal} {Phys. Rev. D}\ }\textbf {\bibinfo {volume} {104}},\ \bibinfo
  {pages} {095006} (\bibinfo {year} {2021}{\natexlab{a}})},\ \Eprint
  {https://arxiv.org/abs/2106.11977} {arXiv:2106.11977 [hep-ph]} \BibitemShut
  {NoStop}%
\bibitem [{\citenamefont {Aranda}\ \emph {et~al.}(2021)\citenamefont {Aranda},
  \citenamefont {Hern{\'a}ndez-Otero}, \citenamefont {Hern{\'a}ndez-Sanchez},
  \citenamefont {Keus}, \citenamefont {Moretti}, \citenamefont
  {Rojas-Ciofalo},\ and\ \citenamefont {Shindou}}]{Aranda:2019vda}%
  \BibitemOpen
  \bibfield  {author} {\bibinfo {author} {\bibfnamefont {A.}~\bibnamefont
  {Aranda}}, \bibinfo {author} {\bibfnamefont {D.}~\bibnamefont
  {Hern{\'a}ndez-Otero}}, \bibinfo {author} {\bibfnamefont {J.}~\bibnamefont
  {Hern{\'a}ndez-Sanchez}}, \bibinfo {author} {\bibfnamefont {V.}~\bibnamefont
  {Keus}}, \bibinfo {author} {\bibfnamefont {S.}~\bibnamefont {Moretti}},
  \bibinfo {author} {\bibfnamefont {D.}~\bibnamefont {Rojas-Ciofalo}},\ and\
  \bibinfo {author} {\bibfnamefont {T.}~\bibnamefont {Shindou}},\ }\bibfield
  {title} {\bibinfo {title} {{Z$_3$ symmetric inert ( 2+1 )-Higgs-doublet
  model}},\ }\href {https://doi.org/10.1103/PhysRevD.103.015023} {\bibfield
  {journal} {\bibinfo  {journal} {Phys. Rev. D}\ }\textbf {\bibinfo {volume}
  {103}},\ \bibinfo {pages} {015023} (\bibinfo {year} {2021})},\ \Eprint
  {https://arxiv.org/abs/1907.12470} {arXiv:1907.12470 [hep-ph]} \BibitemShut
  {NoStop}%
\bibitem [{\citenamefont {Kannike}(2012)}]{Kannike:2012pe}%
  \BibitemOpen
  \bibfield  {author} {\bibinfo {author} {\bibfnamefont {K.}~\bibnamefont
  {Kannike}},\ }\bibfield  {title} {\bibinfo {title} {{Vacuum Stability
  Conditions From Copositivity Criteria}},\ }\href
  {https://doi.org/10.1140/epjc/s10052-012-2093-z} {\bibfield  {journal}
  {\bibinfo  {journal} {Eur. Phys. J. C}\ }\textbf {\bibinfo {volume} {72}},\
  \bibinfo {pages} {2093} (\bibinfo {year} {2012})},\ \Eprint
  {https://arxiv.org/abs/1205.3781} {arXiv:1205.3781 [hep-ph]} \BibitemShut
  {NoStop}%
\bibitem [{\citenamefont {Faro}\ and\ \citenamefont
  {Ivanov}(2019)}]{Faro:2019vcd}%
  \BibitemOpen
  \bibfield  {author} {\bibinfo {author} {\bibfnamefont {F.~S.}\ \bibnamefont
  {Faro}}\ and\ \bibinfo {author} {\bibfnamefont {I.~P.}\ \bibnamefont
  {Ivanov}},\ }\bibfield  {title} {\bibinfo {title} {{Boundedness from below in
  the $U(1)\times U(1)$ three-Higgs-doublet model}},\ }\href
  {https://doi.org/10.1103/PhysRevD.100.035038} {\bibfield  {journal} {\bibinfo
   {journal} {Phys. Rev. D}\ }\textbf {\bibinfo {volume} {100}},\ \bibinfo
  {pages} {035038} (\bibinfo {year} {2019})},\ \Eprint
  {https://arxiv.org/abs/1907.01963} {arXiv:1907.01963 [hep-ph]} \BibitemShut
  {NoStop}%
\bibitem [{\citenamefont {Bento}\ \emph {et~al.}(2022)\citenamefont {Bento},
  \citenamefont {Romão},\ and\ \citenamefont {Silva}}]{Bento_2022}%
  \BibitemOpen
  \bibfield  {author} {\bibinfo {author} {\bibfnamefont {M.~P.}\ \bibnamefont
  {Bento}}, \bibinfo {author} {\bibfnamefont {J.~C.}\ \bibnamefont {Romão}},\
  and\ \bibinfo {author} {\bibfnamefont {J.~P.}\ \bibnamefont {Silva}},\
  }\bibfield  {title} {\bibinfo {title} {Unitarity bounds for all
  symmetry-constrained 3hdms},\ }\bibfield  {journal} {\bibinfo  {journal}
  {Journal of High Energy Physics}\ }\textbf {\bibinfo {volume} {2022}},\ \href
  {https://doi.org/10.1007/jhep08(2022)273} {10.1007/jhep08(2022)273} (\bibinfo
  {year} {2022})\BibitemShut {NoStop}%
\bibitem [{\citenamefont {Boto}\ \emph
  {et~al.}(2021{\natexlab{b}})\citenamefont {Boto}, \citenamefont {Romão},\
  and\ \citenamefont {Silva}}]{Boto_2021}%
  \BibitemOpen
  \bibfield  {author} {\bibinfo {author} {\bibfnamefont {R.}~\bibnamefont
  {Boto}}, \bibinfo {author} {\bibfnamefont {J.~C.}\ \bibnamefont {Romão}},\
  and\ \bibinfo {author} {\bibfnamefont {J.~P.}\ \bibnamefont {Silva}},\
  }\bibfield  {title} {\bibinfo {title} {Current bounds on the type-z $z_3$
  three-higgs-doublet model},\ }\bibfield  {journal} {\bibinfo  {journal}
  {Physical Review D}\ }\textbf {\bibinfo {volume} {104}},\ \href
  {https://doi.org/10.1103/physrevd.104.095006} {10.1103/physrevd.104.095006}
  (\bibinfo {year} {2021}{\natexlab{b}})\BibitemShut {NoStop}%
\bibitem [{\citenamefont {Bahl}\ \emph {et~al.}(2023)\citenamefont {Bahl},
  \citenamefont {Biekötter}, \citenamefont {Heinemeyer}, \citenamefont {Li},
  \citenamefont {Paasch}, \citenamefont {Weiglein},\ and\ \citenamefont
  {Wittbrodt}}]{Bahl_2023}%
  \BibitemOpen
  \bibfield  {author} {\bibinfo {author} {\bibfnamefont {H.}~\bibnamefont
  {Bahl}}, \bibinfo {author} {\bibfnamefont {T.}~\bibnamefont {Biekötter}},
  \bibinfo {author} {\bibfnamefont {S.}~\bibnamefont {Heinemeyer}}, \bibinfo
  {author} {\bibfnamefont {C.}~\bibnamefont {Li}}, \bibinfo {author}
  {\bibfnamefont {S.}~\bibnamefont {Paasch}}, \bibinfo {author} {\bibfnamefont
  {G.}~\bibnamefont {Weiglein}},\ and\ \bibinfo {author} {\bibfnamefont
  {J.}~\bibnamefont {Wittbrodt}},\ }\bibfield  {title} {\bibinfo {title}
  {Higgstools: Bsm scalar phenomenology with new versions of higgsbounds and
  higgssignals},\ }\href {https://doi.org/10.1016/j.cpc.2023.108803} {\bibfield
   {journal} {\bibinfo  {journal} {Computer Physics Communications}\ }\textbf
  {\bibinfo {volume} {291}},\ \bibinfo {pages} {108803} (\bibinfo {year}
  {2023})}\BibitemShut {NoStop}%
\bibitem [{\citenamefont {Borzumati}\ and\ \citenamefont
  {Greub}(1998)}]{Borzumati_1998}%
  \BibitemOpen
  \bibfield  {author} {\bibinfo {author} {\bibfnamefont {F.~M.}\ \bibnamefont
  {Borzumati}}\ and\ \bibinfo {author} {\bibfnamefont {C.}~\bibnamefont
  {Greub}},\ }\bibfield  {title} {\bibinfo {title} {Two higgs doublet model
  predictions for $b \rightarrow x_s \gamma$ in nlo qcd},\ }\bibfield
  {journal} {\bibinfo  {journal} {Physical Review D}\ }\textbf {\bibinfo
  {volume} {58}},\ \href {https://doi.org/10.1103/physrevd.58.074004}
  {10.1103/physrevd.58.074004} (\bibinfo {year} {1998})\BibitemShut {NoStop}%
\bibitem [{\citenamefont {Akeroyd}\ \emph {et~al.}(2021)\citenamefont
  {Akeroyd}, \citenamefont {Moretti}, \citenamefont {Shindou},\ and\
  \citenamefont {Song}}]{Akeroyd_2021}%
  \BibitemOpen
  \bibfield  {author} {\bibinfo {author} {\bibfnamefont {A.}~\bibnamefont
  {Akeroyd}}, \bibinfo {author} {\bibfnamefont {S.}~\bibnamefont {Moretti}},
  \bibinfo {author} {\bibfnamefont {T.}~\bibnamefont {Shindou}},\ and\ \bibinfo
  {author} {\bibfnamefont {M.}~\bibnamefont {Song}},\ }\bibfield  {title}
  {\bibinfo {title} {Cp asymmetries of $\bar{B} \rightarrow x_s/x_d \gamma$ in
  models with three higgs doublets},\ }\bibfield  {journal} {\bibinfo
  {journal} {Physical Review D}\ }\textbf {\bibinfo {volume} {103}},\ \href
  {https://doi.org/10.1103/physrevd.103.015035} {10.1103/physrevd.103.015035}
  (\bibinfo {year} {2021})\BibitemShut {NoStop}%
\bibitem [{\citenamefont {Ellis}\ \emph {et~al.}(2011)\citenamefont {Ellis},
  \citenamefont {Stirling},\ and\ \citenamefont {Webber}}]{Ellis:1996mzs}%
  \BibitemOpen
  \bibfield  {author} {\bibinfo {author} {\bibfnamefont {R.~K.}\ \bibnamefont
  {Ellis}}, \bibinfo {author} {\bibfnamefont {W.~J.}\ \bibnamefont
  {Stirling}},\ and\ \bibinfo {author} {\bibfnamefont {B.~R.}\ \bibnamefont
  {Webber}},\ }\href {https://doi.org/10.1017/CBO9780511628788} {\emph
  {\bibinfo {title} {{QCD and collider physics}}}},\ Vol.~\bibinfo {volume}
  {8}\ (\bibinfo  {publisher} {Cambridge University Press},\ \bibinfo {year}
  {2011})\BibitemShut {NoStop}%
\bibitem [{\citenamefont {Aryshev}\ \emph {et~al.}(2022)\citenamefont {Aryshev}
  \emph {et~al.}}]{ILCInternationalDevelopmentTeam:2022izu}%
  \BibitemOpen
  \bibfield  {author} {\bibinfo {author} {\bibfnamefont {A.}~\bibnamefont
  {Aryshev}} \emph {et~al.} (\bibinfo {collaboration} {ILC International
  Development Team}),\ }\bibfield  {title} {\bibinfo {title} {{The
  International Linear Collider: Report to Snowmass 2021}},\ }\href@noop {} {\
  (\bibinfo {year} {2022})},\ \Eprint {https://arxiv.org/abs/2203.07622}
  {arXiv:2203.07622 [physics.acc-ph]} \BibitemShut {NoStop}%
\bibitem [{ILC(2013)}]{ILC:2013jhg}%
  \BibitemOpen
  \bibfield  {title} {\bibinfo {title} {{The International Linear Collider
  Technical Design Report - Volume 2: Physics}},\ }\href@noop {} {\  (\bibinfo
  {year} {2013})},\ \Eprint {https://arxiv.org/abs/1306.6352} {arXiv:1306.6352
  [hep-ph]} \BibitemShut {NoStop}%
\bibitem [{\citenamefont {Badea}\ \emph {et~al.}(2022)\citenamefont {Badea},
  \citenamefont {Fawcett}, \citenamefont {Huth}, \citenamefont {Khoo},
  \citenamefont {Poggi},\ and\ \citenamefont {Lee}}]{Badea:2022dzb}%
  \BibitemOpen
  \bibfield  {author} {\bibinfo {author} {\bibfnamefont {A.}~\bibnamefont
  {Badea}}, \bibinfo {author} {\bibfnamefont {W.~J.}\ \bibnamefont {Fawcett}},
  \bibinfo {author} {\bibfnamefont {J.}~\bibnamefont {Huth}}, \bibinfo {author}
  {\bibfnamefont {T.~J.}\ \bibnamefont {Khoo}}, \bibinfo {author}
  {\bibfnamefont {R.}~\bibnamefont {Poggi}},\ and\ \bibinfo {author}
  {\bibfnamefont {L.}~\bibnamefont {Lee}},\ }\bibfield  {title} {\bibinfo
  {title} {{Solving combinatorial problems at particle colliders using machine
  learning}},\ }\href {https://doi.org/10.1103/PhysRevD.106.016001} {\bibfield
  {journal} {\bibinfo  {journal} {Phys. Rev. D}\ }\textbf {\bibinfo {volume}
  {106}},\ \bibinfo {pages} {016001} (\bibinfo {year} {2022})},\ \Eprint
  {https://arxiv.org/abs/2201.02205} {arXiv:2201.02205 [hep-ph]} \BibitemShut
  {NoStop}%
\bibitem [{\citenamefont {Kim}\ \emph {et~al.}(2021)\citenamefont {Kim},
  \citenamefont {Ko}, \citenamefont {Park},\ and\ \citenamefont
  {Park}}]{Kim:2021wrr}%
  \BibitemOpen
  \bibfield  {author} {\bibinfo {author} {\bibfnamefont {M.}~\bibnamefont
  {Kim}}, \bibinfo {author} {\bibfnamefont {P.}~\bibnamefont {Ko}}, \bibinfo
  {author} {\bibfnamefont {J.-h.}\ \bibnamefont {Park}},\ and\ \bibinfo
  {author} {\bibfnamefont {M.}~\bibnamefont {Park}},\ }\bibfield  {title}
  {\bibinfo {title} {{Leveraging Quantum Annealer to identify an Event-topology
  at High Energy Colliders}},\ }\href@noop {} {\  (\bibinfo {year} {2021})},\
  \Eprint {https://arxiv.org/abs/2111.07806} {arXiv:2111.07806 [hep-ph]}
  \BibitemShut {NoStop}%
\end{thebibliography}%

\end{document}